\documentclass[pra,reprint]{revtex4-1}
\usepackage{graphicx}
\usepackage{appendix}
\usepackage{braket}
\usepackage{rotating}
\usepackage{simplewick}
\usepackage{amssymb}
\usepackage{amsmath}
\usepackage{txfonts}
\usepackage{bm}
\usepackage{color}
\usepackage{multirow}
\usepackage{booktabs}
\usepackage{dcolumn}

\begin{document}

\title{Nonconvergence of the Feynman--Dyson diagrammatic perturbation expansion of propagators}

\author{So \surname{Hirata}}
\email{sohirata@illinois.edu}
\affiliation{Department of Chemistry, University of Illinois at Urbana-Champaign, Urbana, Illinois 61801, USA}

\author{Ireneusz \surname{Grabowski}}
\affiliation{Institute of Physics, Faculty of Physics, Astronomy, and Informatics, Nicolaus Copernicus University in Toru\'n,
ul.~Grudzi\c{a}dzka 5, 87-100 Toru\'n, Poland}

\author{J. V. \surname{Ortiz}}
\affiliation{Department of Chemistry and Biochemistry, Auburn University, Auburn, Alabama 36849-5312, USA}

\author{Rodney J. \surname{Bartlett}}
\affiliation{Quantum Theory Project, Departments of Chemistry and Physics, University of Florida, Gainesville, Florida 32611, USA}

\date{\today}

\begin{abstract}
Using a general-order {\it ab initio} many-body Green's function method, we numerically illustrate several pathological behaviors of
the Feynman--Dyson diagrammatic perturbation expansion of  one-particle many-body Green's functions as electron Feynman propagators.
(i) The perturbation expansion of the frequency-dependent self-energy
is not convergent at the exact self-energy in many frequency domains. 
(ii) An odd-perturbation-order self-energy
has a qualitatively wrong shape and, as a result, many roots of the corresponding Dyson equation are nonphysical in that 
the poles may be complex or residues can exceed unity or be negative. 
(iii) A higher even-order self-energy consists of vertical lines at many frequencies, predicting numerous phantom poles with zero residues.
(iv) Infinite partial resummations of diagrams by vertex or edge renormalization tend to exacerbate these pathologies.
(v) The nonconvergence is caused by the nonanalyticity of the rational-function form of the exact Green's function at many frequencies, 
where the radius of convergence of its Taylor expansion is zero. This is consistent with the fact that (vi) Pad\'{e} approximants (power-series expansions of a rational function)
can largely restore the correct shape and poles of the Green's function.
Nevertheless, not only does the nonconvergence render higher-order Feynman--Dyson diagrammatic perturbation
theory useless for many lower-lying ionization or higher-lying electron-attachment states, but 
it also calls into question the validity of its combined use with the ans\"{a}tze requiring the knowledge of all poles and residues. 
Such ans\"{a}tze include the Galitskii--Migdal identity, the self-consistent Green's function methods, and some models of the algebraic diagrammatic construction.
\end{abstract}

\maketitle 

\section{Introduction}

In an influential paper \cite{Dyson1952}, Dyson argued that the Feynman--Dyson diagrammatic perturbation theory for quantum electrodynamics is inherently divergent 
in the presence of electron-positron pair formations even after mass and charge are renormalized.
In another important paper \cite{kohn}, Kohn and Luttinger predicted that the finite-temperature diagrammatic perturbation theory for electrons \cite{JhaHirata,HirataJha,HirataJha2,JhaHirata_canonical,HirataJCP}
does not necessarily reduce to the zero-temperature counterpart as the temperature is lowered to zero. 
In particular, it was argued \cite{kohn,luttingerward} that the second-order correction to the grand potential can be divergent when the zeroth-order wave function is degenerate, 
whereas the same correction is finite at zero temperature. 
These predictions have been confirmed both analytically and numerically \cite{hiratapra,HirataCPL}. 
Other instances of breakdown of diagrammatic perturbation theory have also been reported \cite{Lani2012,Schafer2013,Jani2014,Stan2015,Kozik2015,Rossi2015,Schafer2016,Tarantino2017,Gunnarsson2017}.

In this article, we reveal and analyze several additional pathological behaviors of the Feynman--Dyson diagrammatic perturbation expansions 
of one-particle many-body Green's functions (MBGF) or Feynman propagators \cite{Feynman,dyson2,DysonS1949,schwinger,sakurai,march,mattuck1992guide,dyson_physicsworld,Fetter,Kadanoff_book}. 
Our analysis is based on the {\it ab initio} electron propagators for molecules \cite{Linderberg65,Hedin,Ohrn65,Linderberg67,GoscinskiLukman,Doll,pickup,simons_3rd,Yaris,linderbergohrn,Freed74,paldus,Ceder,cederbaumacp,simonsrev,herman,Herman1980,Herman1980_2,BakerPickup,ohrnborn,jorgensensimons,schirmer1982, schirmer,vonniessen,Prasad,GWLouie,oddershede,Kutzelnigg,GW1,ortiz_aqc,GW2,Wire,willow_dyson2,deltamp,Hirata2017,Johnson2018,Doran2019,Opoku2021,Ortiz2022,Opoku2022,Opoku2023,Ortiz2023} as we can take advantage of several independent methods that can determine 
the poles and residues of their exact (finite-basis-set) Green's functions \cite{hirata_ipeomcc} as well as the algorithm that can evaluate perturbation corrections to their frequency-dependent 
self-energy and Green's function at any arbitrary order and frequency \cite{Hirata2017}. 
The conclusions drawn here, however, should be valid for other systems that are studied by  the same theory, such as anharmonic molecular vibrations \cite{Hermes2013mp2}, 
anharmonic lattice vibrations \cite{Qin2020}, energy bands in solids \cite{suhai_qp,sun_qp,SunBartlett1996,Hirata_GF2,hirata_qp,willow_qp,Hirata2023}, 
 finite nuclei, and nuclear matter \cite{Day1967,Rowe1968,Baldo_book}.

Specifically, we show that (i) the perturbation expansion of the frequency-dependent self-energy is nonconvergent at the exact self-energy in many domains of frequency. 
(ii) An odd-perturbation-order self-energy has a qualitatively wrong shape except in the central domain that encloses zero frequency and most principal roots. 
While the diagonal exact self-energy is monotonically decreasing within each frequency bracket separated by its singularities, 
the diagonal odd-order self-energy can be convex, concave, or monotonically increasing in each bracket.
As a result, many roots of the corresponding Dyson equation can be complex and thus nonphysical; when they are real, 
the corresponding residue can fall outside of the valid range of zero to one. 
(iii) An even-order self-energy may have a qualitatively correct, monotonically decreasing form within each bracket demarcated by its singularities.
However, they become more and more vertical with increasing perturbation order, predicting numerous ``phantom'' poles with zero residues. 

None of these problems is detected in the exact self-energy or Green's function, which can be determined (within a finite basis set)  
by the full configuration-interaction (FCI) or, equivalently, by the full equation-of-motion coupled-cluster (EOM-CC) method \cite{Hirata2017,hirata_ipeomcc},
which may be viewed as an infinite resummation of the Feynman--Dyson perturbation corrections \cite{deltamp}.
However, (iv) 
other infinite partial resummations of propagator diagrams by vertex or edge renormalization such as 
the Tamm--Dancoff approximation \cite{Day1967,Kadanoff_book,Linderberg67,PurvisOhrn1974,Schirmer1978,Walter1981,Baldo_book} and 
self-consistent second-order Green's function method \cite{luttingerward,BaymKadanoff1961,Baym_selfconsistent,VanNeck1991,Dickhoff_chapter7,Dickhoff2004,Dahlen2005,Barbieri2006,Barbieri2009,Phillips2014,Neuhauser2017,Tarantino2017,CoveneyTew2023} are shown to 
exhibit the same pathologies with even greater severity.

We also elucidate (v) the cause of the nonconvergence to be the fact that the definition of the Green's function is nonanalytic
at many frequencies, where the radius of convergence of its Taylor expansions is zero. 
In other words, the breakdown of the Feynman--Dyson perturbation theory 
occurs due to a purely mathematical reason, i.e., that the rational-function form 
of the exact Green's function does not lend itself to a converging Taylor expansion, even though physical information such as 
the converging perturbation expansions of all state energies are actually contained in it. 
Consequently, (vi) the general mathematical technique of Pad\'{e} approximants \cite{Goscinski1967,GoscinskiLukman,Brandas1970,Linderberg1970,GoscinskiBrandas1971,Bartlett1972,Bartlett1973,laidig,Bender,hirata_cc},
which are power-series expansions of a rational function, can 
extract this physical information, systematically and rapidly restoring the correct
shape and poles of the Green's function.

Nevertheless, the nonconvergence of the Feynman--Dyson perturbation theory, despite being the mathematical foundation of quantum field theory \cite{Feynman,dyson2,DysonS1949,schwinger,sakurai,dyson_physicsworld}, 
poses difficulties when applying higher-than-second-order approximations to the Green's function methods that are predicated on the knowledge 
of all poles and residues. Such methods include the Galitskii--Migdal identity \cite{Galitskii1958,Koltun1972,Holleboom1990,Ortiz_contour1995}, self-consistent Green's function methods \cite{luttingerward,BaymKadanoff1961,Baym_selfconsistent,VanNeck1991,Dickhoff_chapter7,Dickhoff2004,Dahlen2005,Barbieri2006,Barbieri2009,Phillips2014,Neuhauser2017,Tarantino2017,CoveneyTew2023}, 
and some models of the algebraic diagrammatic construction (ADC) \cite{schirmer1982,schirmer,Deleuze1995,Deleuze_sizeconsistency}. 
The Luttinger--Ward functional \cite{luttingerward,BaymKadanoff1961,Baym_selfconsistent,Kozik2015,Rossi2015,Gunnarsson2017,Lin2018}, which serves as a basis of the self-consistent Green's function methods and 
dynamical mean-field theory (DMFT) \cite{Georges1996}, may also be negatively impacted by these failures. 

\section{Exact Feynman propagator\label{sec:exact}}

\subsection{Formalisms}

An electron Feynman propagator is defined in the time ($t$) domain as a time-ordered sum of Green's functions, 
\begin{eqnarray}
G_{pq}(t) &=& i\theta(-t) \langle \Psi_0 | \hat{p}^\dagger \exp\{i(\hat{H}-E_0)t\} \,\hat{q} | \Psi_0 \rangle
\nonumber\\
&& - i\theta(t) \langle \Psi_0 | \hat{q} \, \exp\{-i(\hat{H}-E_0)t\}\, \hat{p}^\dagger | \Psi_0 \rangle, \label{eq:G}
\end{eqnarray}
where $\theta(t)$ is the Heaviside step function, $\Psi_0$ and $E_0$ are the exact wave function and energy for the $N$ electron ground state,
and $\hat{p}^\dagger$ and $\hat{q}$ are the electron creation and annihilation operators. It describes the probability of an electron (hole)
traveling from the $p$th ($q$th) to $q$th ($p$th) spinorbital in time $t$ ($-t$). 

A Fourier transform of Eq.\ (\ref{eq:G}) yields the electron propagator in the frequency ($\omega$) domain,
\begin{eqnarray}
G_{pq}(\omega) &=& \sum_{I}^{\text{IP}} \frac{\langle \Psi_{0}|\hat{p}^\dagger|\Psi_{I}\rangle \langle \Psi_{I}|\hat{q}|\Psi_{0}\rangle}{\omega - E_{0} + E_{I} - i\eta} 
\nonumber\\&& 
+ \sum_{A}^{\text{EA}} \frac{\langle \Psi_{0}|\hat{q} |\Psi_{A}\rangle \langle \Psi_{A}|\hat{p}^\dagger|\Psi_{0}\rangle}{\omega - E_{A} + E_{0} + i\eta} , \label{eq:Lehman}
\end{eqnarray}
where $\eta$ is a positive infinitesimal, $I$ sums over all $N-1$ electron exact states, and $A$ runs over all $N+1$ electron exact states. 
(Here, $\eta$ is just mathematical convenience and has nothing to do with a lifetime of spectral bandwidth, which are infinite and zero, respectively.
Nonetheless, the present formulation can be applied to solids without any modification, in which 
a spectral bandwidth emerges naturally and rigorously as a manifold of closely packed $\delta$-function peaks.) 

The first term diverges whenever $\omega$ coincides with an exact ionization potential (IP), 
whereas the second term has a pole at an exact electron-attachment energy (EA), apart from their signs.
The primary utility of the electron propagator for molecules and solids 
is the direct determination of IPs and EAs (or electron-correlated energy bands) for both principal (Koopmans) and satellite (shakeup or non-Koopmans) states.
It should be noted, however, the distinction between the principal and satellite roots is not definite, and we use these labels rather loosely in this article. 

\begin{figure}
  \includegraphics[scale=0.3]{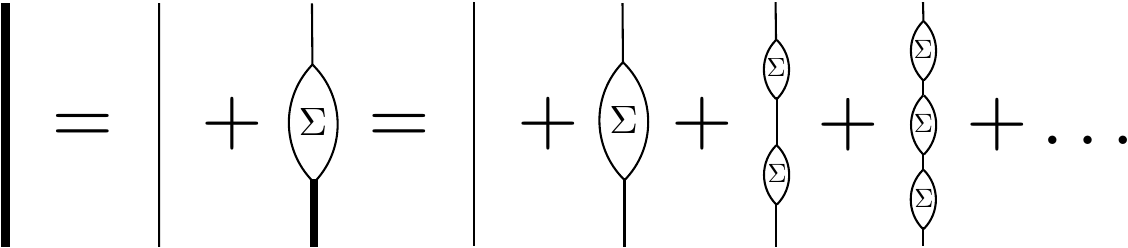}
\caption{The Dyson equations [Eqs.\ (\ref{eq:shortDyson}) and (\ref{eq:longDyson})]. A bold line denotes the exact Green's function, while a thin line designates a zeroth-order Green's function.
A marquise labeled $\Sigma$ contains a complex diagram structure of the exact irreducible self-energy.}
\label{fig:Dyson}
\end{figure}

The exact self-energy $\bm{\Sigma}(\omega)$ is defined by the Dyson equations,
\begin{eqnarray}
\bm{G}(\omega) &=& \bm{G}^{(0)}(\omega) + \bm{G}^{(0)}(\omega) \bm{\Sigma}(\omega) \bm{G}(\omega) \label{eq:shortDyson}\\
&=& \bm{G}^{(0)}(\omega) + \bm{G}^{(0)}(\omega) \bm{\Sigma}(\omega)\bm{G}^{(0)}(\omega) \nonumber\\
&&+ \bm{G}^{(0)}(\omega) \bm{\Sigma}(\omega)\bm{G}^{(0)}(\omega) \bm{\Sigma}(\omega)\bm{G}^{(0)}(\omega) + \dots, \label{eq:longDyson}
\end{eqnarray}
with the zeroth-order Green's function given by
\begin{eqnarray}
G^{(0)}_{pq}(\omega) &=& \sum_{i}^{\text{occ.}} \frac{\delta_{pi}\delta_{qi}}{\omega - \epsilon_i - i\eta} 
+ \sum_{a}^{\text{vir.}} \frac{\delta_{pa}\delta_{qa}}{\omega - \epsilon_a + i\eta}, \label{eq:G0}
\end{eqnarray}
where `occ.'\ and `vir.'\ stand for occupied and virtual spinorbitals in the $N$ electron ground state of a mean-field theory such as the Hartree--Fock (HF) theory,
and $\epsilon_p$ denotes the $p$th canonical spinorbital energy. Throughout this article, we adhere to the convention \cite{shavitt} that $i$, $j$, $k$, and $l$ label occupied spinorbitals,
$a$, $b$, $c$, and $d$ virtual spinorbitals, and $p$ and $q$ either.
$\bm{G}(\omega)$, $\bm{G}^{(0)}(\omega)$ and $\bm{\Sigma}(\omega)$ are $m$-by-$m$ Hermitian matrices with $m$ being the number of spinorbitals.
A diagrammatic representation of the Dyson equations is given in Fig.\ \ref{fig:Dyson}. 

One can formally solve Eq.\ (\ref{eq:shortDyson}) for $\bm{\Sigma}(\omega)$,
\begin{eqnarray}
\bm{\Sigma}(\omega) &=& \Big\{ \bm{G}^{(0)}(\omega) \Big\}^{-1} - \Big\{ \bm{G}(\omega) \Big\}^{-1} \nonumber \\
&=& {\omega}\bm{1}- \bm{\epsilon} - \Big\{ \bm{G}(\omega) \Big\}^{-1}, \label{eq:GF2Sigma}
\end{eqnarray}
which can be inverted to yield
\begin{eqnarray}
\bm{G}(\omega)&=& \Big\{ {\omega}\bm{1} - \bm{\epsilon} - \bm{\Sigma}(\omega) \Big\}^{-1}, \label{eq:Sigma2GF}
\end{eqnarray}
where $\bm{1}$ and $\bm{\epsilon}$ are, respectively, the unit matrix and the diagonal matrix of $\epsilon_p$, which are of the same rank as $\bm{G}(\omega)$ or $\bm{\Sigma}(\omega)$.

One can therefore determine all poles and residues of $\bm{G}(\omega)$ by solving
\begin{eqnarray}
 \Big| {\omega}\bm{1} - \bm{\epsilon} - \bm{\Sigma}(\omega) \Big| = 0, \label{eq:determinant}
\end{eqnarray}
for $\omega$, which are, in turn, roots of the eigenvalue equation,
\begin{eqnarray}
\Big\{ \bm{\epsilon} + \bm{\Sigma}(\omega_q) \Big\} \bm{U}_q =  \bm{U}_q \omega_q, \label{eq:inverseDyson}
\end{eqnarray}
where $\bm{U}_q$ is the $q$th vector of the unitary matrix that brings $\bm{\epsilon} + \bm{\Sigma}(\omega_q)$ into a diagonal form.
The eigenvalue $\omega_q$ reports an exact IP or EA with the corresponding $\bm{U}_q$ defining the so-called Dyson orbital \cite{OrtizDyson}.
This equation, known as the inverse Dyson equation, 
has a striking physical interpretation as {\it an exact one-electron equation} with the nonlocal, frequency-dependent correlation potential $\bm{\Sigma}(\omega)$ \cite{Bartlett2005,bartlett1p}.

The residue $F(\omega_q)$ for the pole $\omega_q$ is evaluated as
\begin{eqnarray}
F(\omega_q) \equiv \text{Res}_{\omega_q} G_{qq}(\omega) = \left\{ 1 - \bm{U}_q^\dagger \left(\frac{\partial \bm{\Sigma}(\omega)}{\partial \omega}\right)_{\omega_q}\bm{U}_q \right\}^{-1}. \label{eq:residues}
\end{eqnarray}
It quantifies a one-electron weight in the many-electron IP or EA state, and is proportional to the transition probability in photoelectron spectroscopy \cite{Manne1970}. 
The residues must therefore add up to the number of electrons ($n_{\text{e}}$)
when summed over all IP poles ($\omega_q < 0$):
\begin{eqnarray}
\sum_{\omega_q}^{\text{IP}} F(\omega_q) = n_{\text{e}}. \label{eq:sumrule1}
\end{eqnarray}

In addition to IPs and EAs, the exact total energy is gleaned from $\bm{G}(\omega)$ with the aid of the
Galitskii--Migdal identity \cite{Galitskii1958,Koltun1972,Holleboom1990,Ortiz_contour1995},
\begin{eqnarray}
E = E_{\text{nuc.}} + \frac{1}{2} \sum_{\omega_q}^{\text{IP}} \left( \bm{U}_q^\dagger \bm{H}^{\text{core}} \bm{U}_q + \omega_q \right) F(\omega_q), \label{eq:GM}
\end{eqnarray}
where the summation is taken over all IP poles ($\omega_q < 0$), $E_{\text{nuc.}}$ is the nuclear repulsion energy, and 
$\bm{H}^{\text{core}}$ is the one-electron part of the Hamiltonian matrix \cite{szabo}. Equation (\ref{eq:GM}) says that the total energy (minus $E_{\text{nuc.}}$) 
is the sum of all IP poles ($\omega_q$) times their one-electron weights (residues) 
corrected for the double counting of the two-electron interactions.

In the diagonal approximation \cite{Hirata2017} to the self-energy, the inverse Dyson equation simplifies to
\begin{eqnarray}
{\epsilon_q} + {\Sigma_{qq}}(\omega_q)  = {\omega_q} . \label{eq:diagonal}
\end{eqnarray}
The residue $F({\omega}_q)$ for the pole ${\omega_q}$ is then computed as
\begin{eqnarray}
F({\omega}_q) \equiv \text{Res}_{\omega_q} G_{qq}(\omega) = \left\{ 1 - \left.\frac{\partial {\Sigma_{qq}}(\omega)}{\partial \omega}\right|_{\omega_q} \right\}^{-1}. \label{eq:residue_diag}
\end{eqnarray}
The sum rule for the residues then becomes
\begin{eqnarray}
\sum_{\omega_q} F({\omega}_q) = 1,\label{eq:sumrule2}
\end{eqnarray}
where the summation is taken over all roots of the $q$th diagonal inverse Dyson equation.

\subsection{Numerical results}

\begin{figure}
  \includegraphics[scale=0.65]{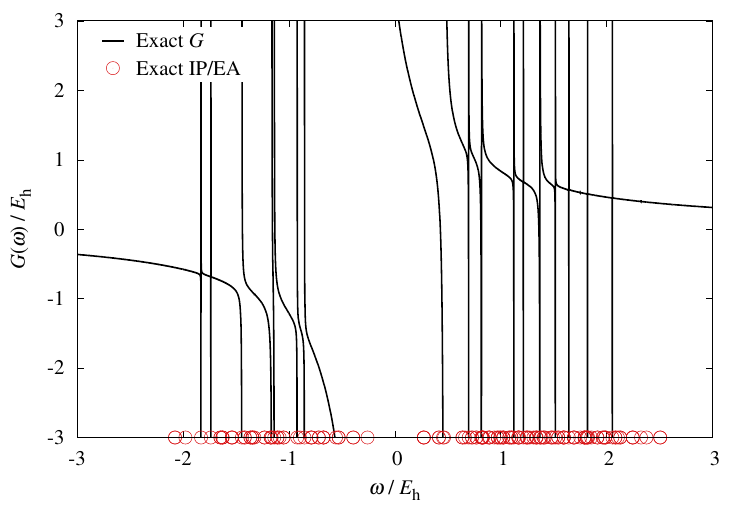}
\caption{The third diagonal element of the exact $\bm{G}(\omega)$ as a function of $\omega$ for the BH molecule (1.232\ \AA\ in the minimal basis set; the third spatial orbital in the increasing order of orbital energy
corresponds to the HOMO). The exact IPs and EAs obtained by FCI are superposed (as open circles), occurring at the poles of the Green's function.
The HF and FCI energies are $-24.752788\,E_{\text{h}}$ and $-24.809940\,E_{\text{h}}$, respectively.}
\label{fig:BH_exactGF}
\end{figure}

In Fig.\ \ref{fig:BH_exactGF} is plotted the third diagonal element of the exact $\bm{G}(\omega)$ matrix as functions of $\omega$ for 
the boron hydride molecule with the bond length of 1.232\ \AA\ in the minimal basis set.
The third element  corresponds to the highest-occupied molecular orbital (HOMO). The other elements of $\bm{G}(\omega)$ are omitted to avoid clutter. 
The exact $\bm{G}(\omega)$ was obtained by literally evaluating Eq.\ (\ref{eq:Lehman}) using a determinant-based FCI program \cite{Hirata2017,deltamp}. 
The figure confirms the well-known fact \cite{Walter_JMathPhys} that the function is divided by singularities into consecutive regions or {\it brackets}, within each of which it
is a monotonically decreasing, ``shoulder-like'' function of $\omega$. 

In the same figure  are superposed the exact IPs and EAs (signs reversed; 300 each) obtained by the determinant-based FCI program \cite{hirata_ipeomcc}. 
They coincide with the poles of $G_{33}(\omega)$ as they should.
There are some IPs and EAs that apparently lack matching poles, but they correspond either to the poles of other elements of $\bm{G}(\omega)$  or 
to nearly vertical poles (with nearly zero residues) that have fallen through the $\omega$ mesh used for plotting Fig.\ \ref{fig:BH_exactGF}. These are merely some minor issues associated
with the algorithms adopted, and there is no pathological behavior detectable in the exact MBGF, which reproduces the results of FCI. 

\begin{figure}
  \includegraphics[scale=0.65]{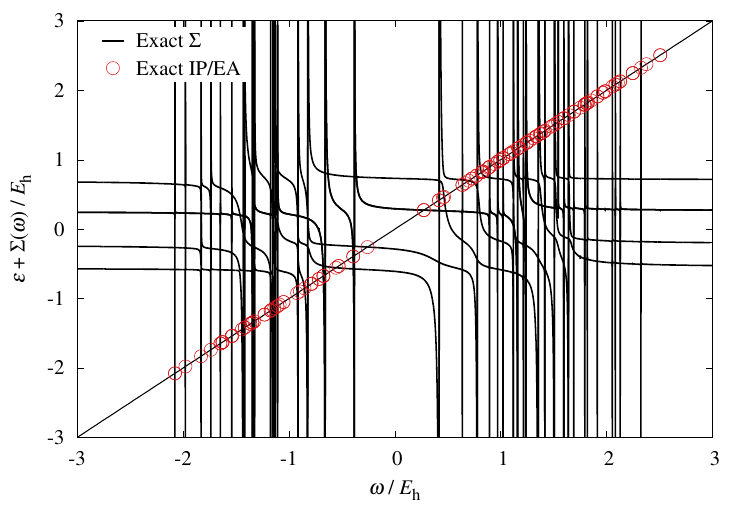}
\caption{Eigenvalues of the exact $\bm{\epsilon}+\bm{\Sigma}(\omega)$ as a function of $\omega$ for the BH molecule.
The exact IPs and EAs obtained by FCI are superposed (as open circles), 
and they coincide with the roots of the inverse Dyson equation, occurring at the intersections of the eigenvalues with the diagonal $\omega$ line; see Eq.\ (\ref{eq:inverseDyson}).}
\label{fig:BH_exact}
\end{figure}

Figure \ref{fig:BH_exact} plots eigenvalues of the exact $\bm{\epsilon} + \bm{\Sigma}(\omega)$ matrix as functions of $\omega$. Of the six eigenvalues 
the lowest one is not visible in this plot and the second and third highest ones are degenerate. 
Each root of the inverse Dyson equation is expected at an intersection of the eigenvalues of $\bm{\epsilon} + \bm{\Sigma}(\omega)$ and the diagonal
line $\omega$ as per Eq.\ (\ref{eq:inverseDyson}). In fact, the IPs and EAs obtained from FCI are seen to occur precisely at these intersections. 
The few IPs and EAs appearing to occur away from any intersection are likely due to the nearly vertical poles with nearly zero residues, which are thus undetected by an $\omega$ mesh. 
This issue may be  a weakness of the graphical method \cite{Walter_JMathPhys,Piecuch1990} of solving the inverse Dyson equation
if one is concerned with determining all  roots, but it by no means signals any fundamental problem in the theory. 

\begin{figure}
 \includegraphics[scale=0.65]{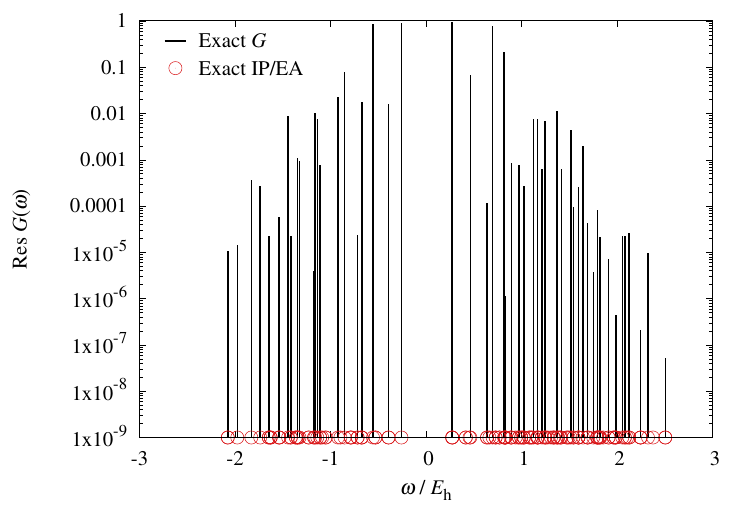}
\caption{Residues at poles $\omega$  of the exact $\bm{G}(\omega)$ for the BH molecule.
The exact IPs and EAs obtained by FCI are superposed (as open circles).}
\label{fig:BH_residues}
\end{figure}

Figure \ref{fig:BH_residues} is a histogram of the poles of the exact $\bm{G}(\omega)$; the height of each impulse 
is the corresponding residue [Eq.\ (\ref{eq:residues})]. There are numerous poles outside of this graph (see Sec.\ \ref{sec:infinite}).
Generally, the poles and residues can be determined by the graphical \cite{Walter_JMathPhys,Piecuch1990}
or arrow-matrix diagonalization method \cite{Walter_JMathPhys}. In this study, we used neither; the poles $\{\omega_q\}$ in their entirety 
were first obtained by  FCI  \cite{hirata_ipeomcc} and verified by substitution to the inverse Dyson equation [Eq.\ (\ref{eq:inverseDyson})].
The residue at $\omega=\omega_q$ was then computed by evaluating Eq.\ (\ref{eq:residues}). The derivative of $\bm{\Sigma}$ with respect to $\omega$
can be taken analytically by
\begin{eqnarray}
\left( \frac{\partial \bm{\Sigma}(\omega)}{\partial \omega}\right)_{\omega_q} &=& \bm{1} + \Big\{\bm{G}(\omega)\Big\}^{-1}\left( \frac{\partial \bm{G}(\omega)}{\partial \omega}\right)_{\omega_q} \Big\{\bm{G}(\omega)\Big\}^{-1} \label{eq:Sigmaderiv}
\end{eqnarray}
with
\begin{eqnarray}
\frac{\partial G_{pq}(\omega)}{\partial \omega} &=& - \sum_{I}^{\text{IP}} \frac{\langle \Psi_{0}|\hat{p}^\dagger|\Psi_{I}\rangle \langle \Psi_{I}|\hat{q}|\Psi_{0}\rangle}{(\omega - E_{0} + E_{I})^2} 
\nonumber\\&& 
- \sum_{A}^{\text{EA}} \frac{\langle \Psi_{0}|\hat{q} |\Psi_{A}\rangle \langle \Psi_{A}|\hat{p}^\dagger|\Psi_{0}\rangle}{(\omega - E_{A} + E_{0})^2} . \label{eq:Gderiv}
\end{eqnarray}
However, the last expression is ill-conditioned at every pole $\omega=\omega_q$. We therefore approximated this derivative as an average
of the derivatives at $\omega = \omega_q \pm 10^{-9}\,E_{\text{h}}$. 

The residues thus obtained correctly fall in the range of zero to one for the BH molecule. 
They also satisfy the particle number sum rule [Eq.\ (\ref{eq:sumrule1})] and 
Galitskii--Migdal identity [Eq.\ (\ref{eq:GM})] with the precision of $10^{-7}$ and $10^{-6}\,E_{\text{h}}$, respectively, which
 is a numerical manifestation of the fact that the exact Green's function obeys the Baym--Kadanoff conservation laws \cite{BaymKadanoff1961,Baym_selfconsistent,Kadanoff_book,Dahlen2005}.
Very many tiny contributions from satellite roots are crucial for these identities to be accurately satisfied, and including only
principal roots results in severe errors even in this tiny system.

Overall, the exact finite-basis-set Green's function and self-energy are well-behaved, satisfying conservation laws and
yielding results that are in exact numerical agreement with alternative methods such as FCI or full EOM-CC. 
Therefore, the pathological behaviors we are about to discuss are exclusively ascribed to their perturbation expansions.

\section{Feynman--Dyson diagrammatic perturbation expansion of propagator}

\subsection{Formalisms}

In most applications, both the Green's function and self-energy 
are expanded in perturbation series. 
In this article, perturbation corrections 
are denoted by symbols prefixed with $\delta$ with its order given as the parenthesized superscript. 
\begin{eqnarray}
\bm{G}(\omega) &=& \bm{G}^{(0)}(\omega) + \delta \bm{G}^{(1)} (\omega)+ \delta \bm{G}^{(2)}(\omega) + \dots, \\
\bm{\Sigma}(\omega) &=& \delta\bm{\Sigma}^{(1)} (\omega)+ \delta\bm{\Sigma}^{(2)}(\omega) + \delta\bm{\Sigma}^{(3)}(\omega) + \dots.
\end{eqnarray}
Cumulative approximations to $\bm{G}$ or $\bm{\Sigma}$ through order $n$ are denoted by symbols without a $\delta$ prefix.
\begin{eqnarray}
\bm{G}^{(n)}(\omega) &\equiv& \bm{G}^{(0)}(\omega) + \sum_{i=1}^n \delta \bm{G}^{(i)}(\omega) , \\
\bm{\Sigma}^{(n)}(\omega) &\equiv& \sum_{i=1}^n \delta\bm{\Sigma}^{(i)} (\omega), 
\end{eqnarray}

In this article, MBGF($n$) refers to the $n$th-order Feynman--Dyson perturbation approximation to one-particle many-body 
Green's function theory, seeking the poles and residues of $\bm{G}^{(n)}$ by solving the inverse Dyson equation with $\bm{\Sigma}^{(n)}$.
On the other hand, the exact MBGF is synonymous to FCI as demonstrated in Sec.\ \ref{sec:exact}. MBGF($n$) is diagrammatically linked \cite{Hirata2017} and 
thus size-consistent \cite{HirataTCA}. Its {\it ab initio} computer implementations based on Gaussian-type orbitals \cite{szabo} can therefore 
be applied to both electrons \cite{Hirata2023} and phonons \cite{Qin2020} in solids. 

The perturbation corrections are usually stipulated diagrammatically \cite{march,mattuck1992guide,dyson_physicsworld,Fetter,Kadanoff_book}, but they
are also defined \cite{Hirata2017} algebraically as
\begin{eqnarray}
\delta \bm{G}^{(n)} (\omega) &=& \frac{1}{n!} \left. \frac{\partial^n \bm{G}(\omega;\lambda)}{\partial \lambda^n}\right|_{\lambda=0}, \\
\delta \bm{\Sigma}^{(n)} (\omega) &=& \frac{1}{n!} \left. \frac{\partial^n \bm{\Sigma}(\omega;\lambda)}{\partial \lambda^n}\right|_{\lambda=0},
\end{eqnarray}
where $\bm{G}(\omega;\lambda)$ and $\bm{\Sigma}(\omega;\lambda)$ are the exact (i.e., FCI) values of the respective quantities 
for a perturbation-scaled Hamiltonian $\hat{H} = \hat{H}^{(0)} + \lambda \hat{V}^{(1)}$. 
The zeroth-order Hamiltonian $\hat{H}^{(0)}$ corresponds to the $\bm{G}^{(0)}$ of Eq.\ (\ref{eq:G0}). 
Here, we adopt the HF theory as the zeroth order, which implies
$\delta{\Sigma}_{pq}^{(1)} = {0}$ \cite{Hirata2017}.

These $\lambda$-derivatives can be taken either numerically
or analytically. From the former, we obtain benchmark data of the perturbation corrections at several low orders \cite{Hirata2017}. From the latter, we derive 
recursions of $\delta \bm{G}^{(n)}$ and $\delta \bm{\Sigma}^{(n)}$ in the style of Rayleigh--Schr\"{o}dinger perturbation theory, which can then be implemented
into a general-order algorithm \cite{Hirata2017}. This strategy was used in this study. The recursions also justify the diagrammatic rules through
the linked-diagram and irreducible-diagram theorems in a time-independent picture \cite{Hirata2017}.

\begin{figure}
  \includegraphics[scale=0.3]{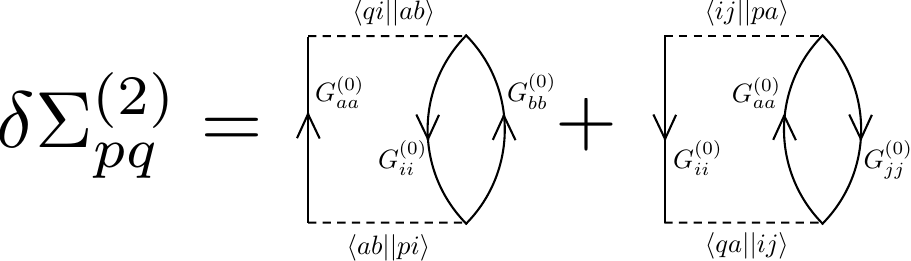}
\caption{The second-order self-energy.}
\label{fig:Sigma2}
\end{figure}

While the time-independent picture is more mathematically transparent and systematically extensible to arbitrarily high orders \cite{Hirata2017}, 
the time-dependent one may be more appealing to our intuition and expedient \cite{dyson_physicsworld,mattuck1992guide}.   
In the latter, for example, the second-order self-energy is stipulated diagrammatically as in  Fig.\ \ref{fig:Sigma2}. It graphically describes the process in which 
(i) a mean-field particle (hole) scatters another particle out of its mean-field state, thereby creating a hole, at one time;
(ii) all three particles and holes propagate in their respective mean-field potentials, i.e., driven by the mean-field propagator $\bm{G}^{(0)}$; and 
(iii) the particle-hole pair recombines at another time. 
The numerical value of $\delta\bm{\Sigma}^{(2)}$, which is related to the probability of this overall process, is the product of the probabilities of the constituent 
scattering and propagation events summed over all possible times and positions of their occurrences. Consulting with Table 4.3 of Mattuck \cite{mattuck1992guide}, we can then evaluate it as
\begin{widetext}
\begin{eqnarray}
\delta \Sigma_{pq}^{(2)}(\omega) 
&=& (-1)^1 i \sum_i^{\text{occ.}} \sum_{a<b}^{\text{vir.}} 
\int_{-\infty}^{\infty}\frac{d\omega_a}{2\pi} 
\int_{-\infty}^{\infty}\frac{d\omega_i}{2\pi} 
\, {(-i)\langle qi||ab \rangle}{(-i)\langle ab|| pi \rangle}   
\,  iG_{aa}^{(0)}(\omega_a) iG_{ii}^{(0)}(\omega_i) iG_{bb}^{(0)}(\omega+\omega_i-\omega_a) \nonumber\\
&& + (-1)^1 i  \sum_a^{\text{vir.}} \sum_{i<j}^{\text{occ.}} 
\int_{-\infty}^{\infty}\frac{d\omega_i}{2\pi} 
\int_{-\infty}^{\infty}\frac{d\omega_a}{2\pi} 
\, (-i) \langle qa|| ij \rangle (-i) \langle ij||pa \rangle 
\,  iG_{ii}^{(0)}(\omega_i) iG_{aa}^{(0)}(\omega_a) iG_{jj}^{(0)}(\omega+\omega_a-\omega_i) \label{eq:contour1}\\
&=& \frac{1}{2} \sum_i^{\text{occ.}} \sum_{a,b}^{\text{vir.}} \frac{\langle qi||ab \rangle\langle ab|| pi \rangle }{\omega + \epsilon_i-\epsilon_a-\epsilon_b} 
+ \frac{1}{2} \sum_{i,j}^{\text{occ.}} \sum_{a}^{\text{vir.}} \frac{\langle qa|| ij \rangle \langle ij||pa \rangle}{\omega + \epsilon_a - \epsilon_i - \epsilon_j},
\label{eq:Sigma2}
\end{eqnarray}
\end{widetext}
where an occupied spinorbital index and the imaginary unit both denoted by ``$i$'' should be distinguished, and $\langle pq||rs \rangle$ is an antisymmetrized two-electron integral \cite{szabo,shavitt}.
Second-order many-body Green's function method [MBGF(2)] solves the inverse Dyson equation [Eq.\ (\ref{eq:inverseDyson})] with this
$\bm{\Sigma}^{(2)}$. Since the roots of this equation occur at the intersection of $\bm\epsilon + \bm\Sigma^{(2)}(\omega)$ and $\omega$, they can never be divergent 
even though the method is perturbative.  

\begin{figure}
  \includegraphics[scale=0.3]{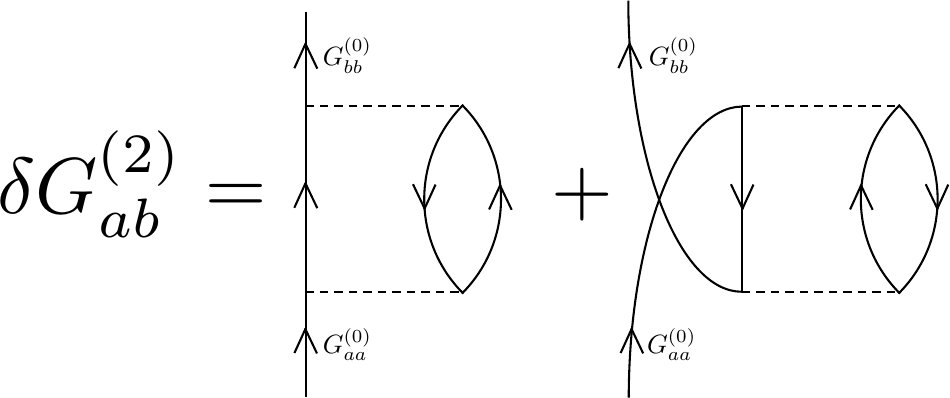}
\caption{The second-order Green's function.}
\label{fig:G2}
\end{figure}

\begin{figure}
  \includegraphics[scale=0.3]{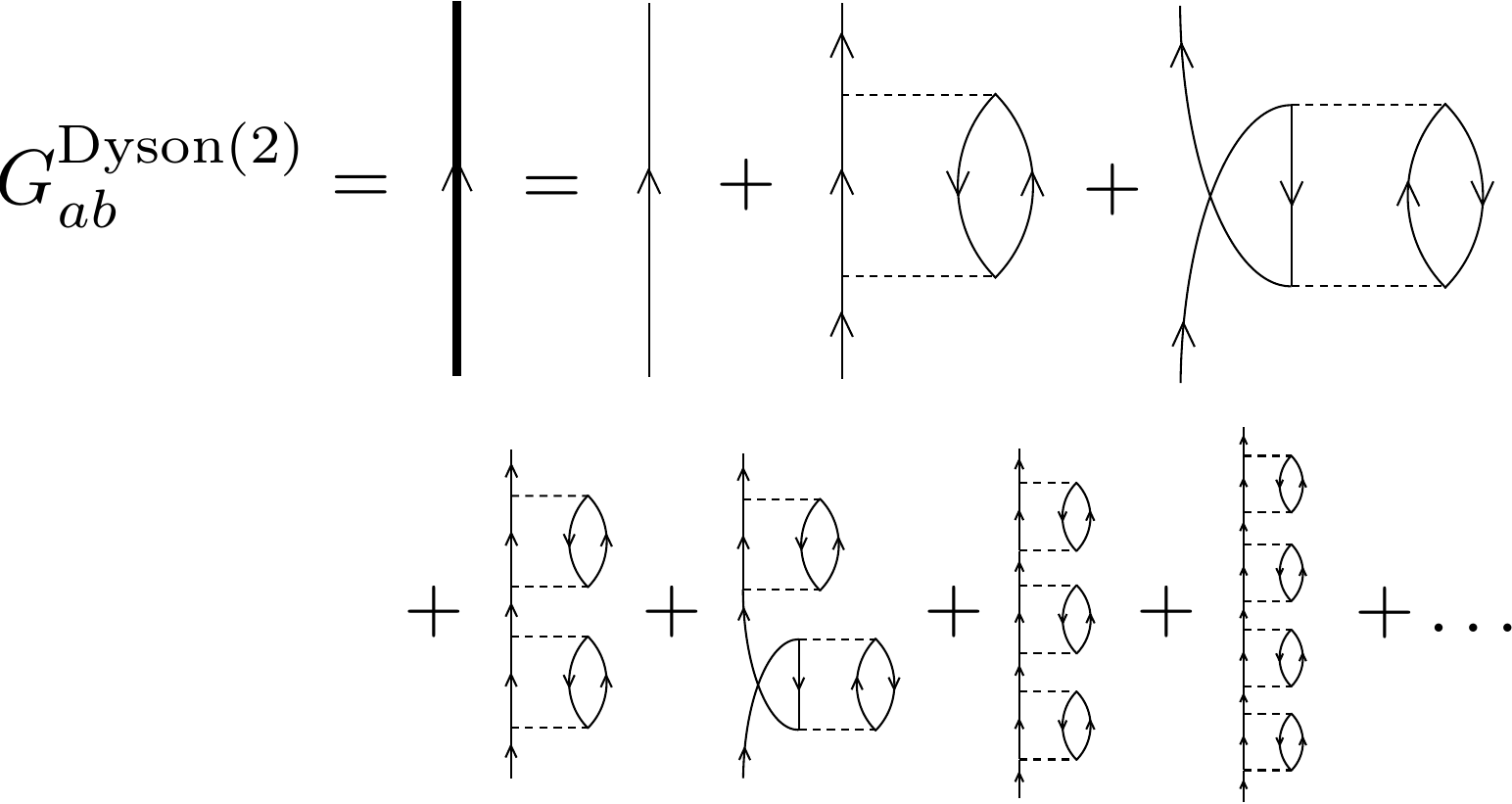}
\caption{The second-order bold-line Green's function.}
\label{fig:GDyson2}
\end{figure}

The second-order correction to the Green's function is then described by the same diagram as the second-order self-energy, but appended with
long dangling edges, as shown in Fig.\ \ref{fig:G2}. It is important to recognize that the roots of the inverse Dyson equation with $\bm{\Sigma}^{(2)}$
are {\it not} the poles of this $\bm{G}^{(2)}$; rather, they are the poles of $\bm{G}^{\text{Dyson}(2)}$  (using the nomenclature of Holleboom and Snijders \cite{Holleboom1990}) defined by
\begin{eqnarray}
\bm{G}^{\text{Dyson}(n)}(\omega) &=& \left\{ \omega\bm{1} - \bm{\epsilon} - \bm{\Sigma}^{(n)}(\omega) \right\}^{-1}, 
\end{eqnarray}
in analogy to Eq.\ (\ref{eq:Sigma2GF}) and in accordance with the Dyson equations [Eqs.\ (\ref{eq:shortDyson}) and (\ref{eq:longDyson})].
Therefore, as shown in Fig.\ \ref{fig:GDyson2}, $\bm{G}^{\text{Dyson}(2)}$ is a {\it bold-line} Green's function (just like the one appearing in Fig.\ \ref{fig:Dyson}), which
includes an infinite-order correction through repeated actions of $\bm\Sigma^{(2)}$. In this sense, MBGF($n$) is an infinite-order theory for IPs and EAs even for a finite perturbation order $n$.

\begin{figure}
  \includegraphics[scale=0.3]{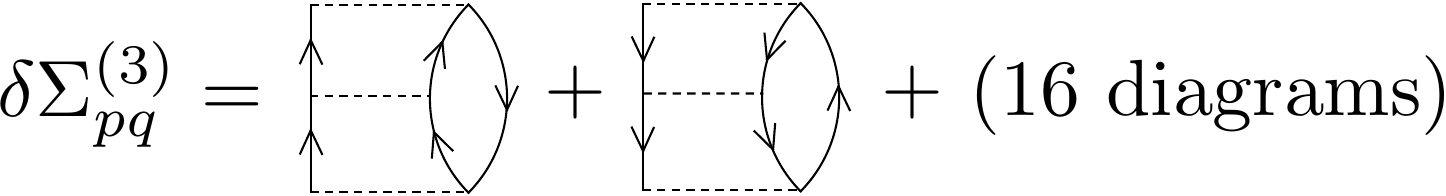}
\caption{The third-order self-energy. See Appendix 1 of \"{O}hrn and Born \cite{ohrnborn} for a complete list.}
\label{fig:Sigma3}
\end{figure}

Likewise, the third-order self-energy is evaluated from its diagrams \cite{ohrnborn} as
\begin{eqnarray}
\delta \Sigma_{pq}^{(3)}(\omega) &=& \frac{1}{4} \sum_{i}^{\text{occ.}} \sum_{a,b,c,d}^{\text{vir.}} \frac{\langle qi || ab \rangle\langle ab || cd \rangle \langle cd || pi \rangle}
{(\omega + \epsilon_i - \epsilon_a - \epsilon_b)(\omega+\epsilon_i - \epsilon_c - \epsilon_d)} \nonumber\\
&& -  \frac{1}{4} \sum_{i,j,k,l}^{\text{occ.}} \sum_{a}^{\text{vir.}} \frac{\langle qa || ij \rangle\langle ij || kl \rangle \langle kl || pa \rangle}
{(\omega + \epsilon_a - \epsilon_i - \epsilon_j)(\omega+\epsilon_a - \epsilon_k - \epsilon_l)} 
\nonumber\\&& 
+ \text{(16 terms)}, \label{eq:Sigma3}
\end{eqnarray}
corresponding to the diagrammatic equation in Fig.\ \ref{fig:Sigma3}. It may be noticed that the functional form of $\delta \bm{\Sigma}^{(3)}$ with respect to $\omega$
is different from that of $\delta \bm{\Sigma}^{(2)}$ [Eq.\ (\ref{eq:Sigma2})]; 
poles in $\delta\bm{\Sigma}^{(3)}$ are second order, while $\delta\bm{\Sigma}^{(2)}$ and the exact Green's function have only first-order poles. 
This difference has a grave consequence on the roots of the inverse Dyson equation, which we now discuss.

\subsection{Numerical results}
 
\begin{figure}
  \includegraphics[scale=0.65]{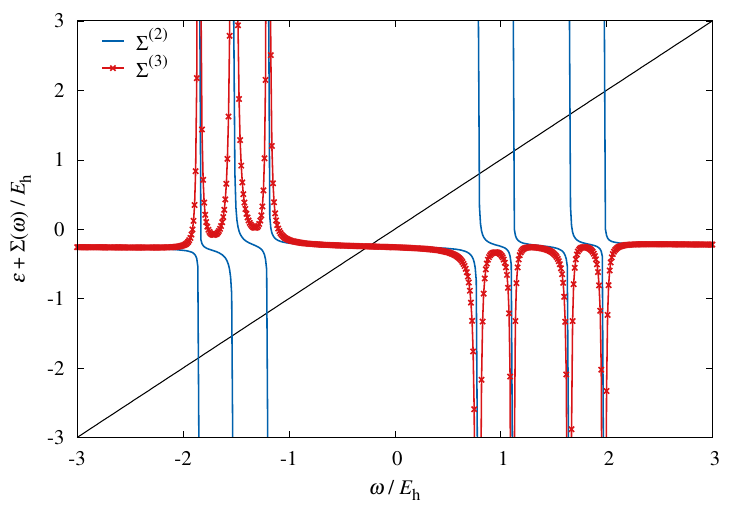}
\caption{The third diagonal elements of $\bm{\epsilon}+\bm{\Sigma}^{(2)}(\omega)$ and $\bm{\epsilon}+\bm{\Sigma}^{(3)}(\omega)$ 
 as functions of $\omega$ for the BH molecule. }
\label{fig:BH_HOMOself23}
\end{figure}

\begin{figure}
  \includegraphics[scale=0.65]{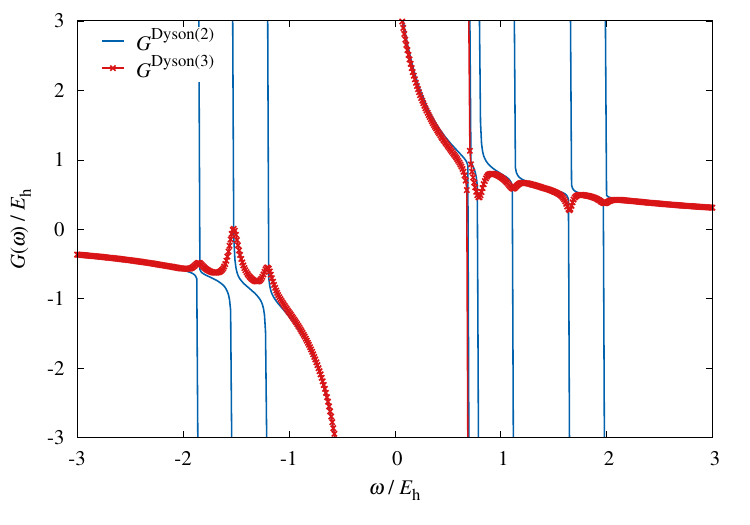}
\caption{The third diagonal elements of $\bm{G}^{\text{Dyson}(2)}(\omega)$ and $\bm{G}^{\text{Dyson}(3)}(\omega)$ as functions of $\omega$ for the BH molecule.}
\label{fig:BH_HOMOGreen23}
\end{figure}

In Fig.\ \ref{fig:BH_HOMOself23} are plotted diagonal elements of the second- and third-order self-energies, $\epsilon_3+\Sigma_{33}^{(2)}$ and $\epsilon_3+\Sigma_{33}^{(3)}$, 
for the third orbital (HOMO) of the BH molecule \cite{Hirata2017}. Intersections of these functions with the diagonal $\omega$ line (also drawn) are the roots of 
the corresponding inverse Dyson equations in the diagonal approximation. See Eq.\ (\ref{eq:diagonal}).

The second-order self-energy has qualitatively the same functional form as the exact self-energy (Fig.\ \ref{fig:BH_exact}) 
in that they are both separated by singularities into consecutive $\omega$ brackets, within each of which they are monotonically decreasing and shoulder-like. 
The singularities of $\bm{\Sigma}^{(2)}$, i.e., the boundaries of the brackets, occur at the two-particle-one-hole (2p1h) 
($\epsilon_a + \epsilon_b - \epsilon_i$) and two-hole-one-particle (2h1p) ($\epsilon_i + \epsilon_j - \epsilon_a$) HF orbital energy differences according to Eq.\ (\ref{eq:Sigma2}). 
In each bracket, a diagonal element of the second-order self-energy intersects the diagonal $\omega$ line exactly once, just as the exact self-energy does \cite{Walter_JMathPhys}. 
The second-order inverse Dyson equation has the principal root for the HOMO at around $-0.25\,E_{\text{h}}$
in the central bracket enclosing $\omega=0$, where the self-energy is relatively flat and whose residue is close to unity [see Eq.\ (\ref{eq:residue_diag})]. 
The self-energy also intersects the $\omega$ line in each of the other brackets, typically on a near-vertical part of its shoulder-like curve.
This means that these satellite roots coincide with the 2p1h or 2h1p 
HF orbital energy differences, accounting for little to no electron-correlation effects, and have near zero residues. 
Nevertheless, MBGF(2) is overall well-behaved. 

The third-order self-energy, in contrast, has a qualitatively wrong functional form. It is still separated by the same 2p1h and 2h1p singularities into the identical set of brackets [see Eq.\ (\ref{eq:Sigma3})],  
but within each bracket, $\bm{\Sigma}^{(3)}$ is either concave or convex except in the central bracket, where it is correctly monotonically decreasing. Consequently,
in the domain of $\omega$ of Fig.\ \ref{fig:BH_HOMOself23}, the third-order inverse Dyson equation has only one real root, i.e., the principal root for the HOMO at around $-0.25\,E_{\text{h}}$. 
The different functional forms of the self-energy between the second and third orders 
can be easily rationalized by their algebraic definitions, Eqs.\ (\ref{eq:Sigma2}) and (\ref{eq:Sigma3}).
The second-order self-energy (as well as the exact self-energy) has only first-order poles, whereas the third-order self-energy features up to second-order poles.  

Figure \ref{fig:BH_HOMOGreen23} plots the second- and third-order bold-line Green's functions.
The $\bm{G}^{\text{Dyson}(2)}$ has the same overall appearance as the exact Green's function (Fig.\ \ref{fig:BH_exactGF}). 
It exhibits a ``fat'' pole at the principal root for the HOMO and several ``thin'' poles at satellite roots. In contrast, $\bm{G}^{\text{Dyson}(3)}$ displays only undulations
but no poles at the frequencies where satellite roots are expected. The diagonal element of $\bm{G}^{\text{Dyson}(3)}$ for the HOMO has poles only at the principal roots 
at around $-0.25\,E_{\text{h}}$ and $0.69\,E_{\text{h}}$, which is consistent with Fig.\ \ref{fig:BH_HOMOself23}. (Although there is no root at around $0.69\,E_{\text{h}}$ in MBGF(3) according to 
Fig.\ \ref{fig:BH_HOMOself23}, this principal EA root has a considerable mixing with other roots with its residue being only $0.77$, which is why it creates a pole in $G_{33}^{\text{Dyson}(3)}$.) 
Hence, the perturbative Green's functions confirm the absence of many real satellite roots at MBGF(3).

 
\begin{figure}
  \includegraphics[scale=0.65]{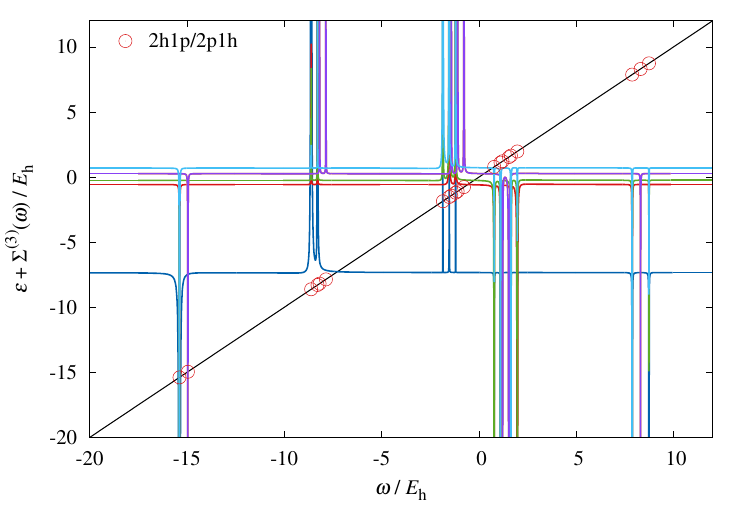}
\caption{All six diagonal elements of $\bm{\epsilon}+\bm\Sigma^{(3)}(\omega)$ 
 as functions of  $\omega$ for the BH molecule. The fourth and fifth elements are degenerate, appearing as one curve. The 2p1h and 2h1p HF orbital energy differences are superposed.}
\label{fig:BH_ALLself3}
\end{figure}

Where have the missing satellite roots of MBGF(3) gone? We believe that they have gone to the complex space, although no attempt 
has been made to determine them numerically. 

Figure \ref{fig:BH_ALLself3} plots all six diagonal elements of $\bm{\epsilon} + \bm{\Sigma}^{(3)}$ in a full domain of $\omega$.
For the HF occupied orbitals 1, 2, 3, and virtual orbital 6 (labeled in the increasing order of energy), which transform as $a_1$ in the $C_{4v}$ subgroup, there are 
twelve (12) 2p1h and 2h1p singularities or thirteen (13) brackets. Hence, the second- and third-order inverse Dyson equations in the diagonal approximation 
are a thirteenth-order polynomial equation for each of these $a_1$  orbitals. The  second-order equation has one principal root and twelve satellite roots (13 in total), all of which are real. 

On the other hand, as Fig.\ \ref{fig:BH_ALLself3} indicates,  
the diagonal element of the third-order self-energy for orbital 1 intersects the $\omega$ line only nine (9) times, meaning that its diagonal third-order inverse Dyson equation has 
at least 9 real roots (some may be degenerate) out of 13 roots in total. The remaining 4 (or less in the case of degeneracy) roots must be complex. 
For orbitals 2, 3, and 6, the number of intersections is three (3) and hence the third-order equation for each of these orbitals has  3 (or more) real and 10 (or less) complex roots. 

The HF virtual orbitals 4 and 5 transform as $e$ of $C_{4v}$ and are degenerate.
There are nine (9) 2p1h and 2h1p singularities or ten (10) brackets for these orbitals. The diagonal second- and third-order inverse Dyson equations are therefore a tenth-order polynomial equation. 
The second-order equation has ten real roots, but as seen in Fig.\ \ref{fig:BH_ALLself3}, the third-order self-energy intersects the $\omega$ line
only three (3) times, implying that the third-order equation has 4 or more real roots and 6 or less complex roots (since complex roots of a polynomial equation with real coefficients 
occur in complex conjugate pairs, at least one of the real roots must be degenerate).

From Fig.\ \ref{fig:BH_ALLself3}, it can furthermore be observed that real satellite roots of MBGF(3) tend to appear as {\it nearly} degenerate pairs or doublets (which is distinct from
the exact degeneracy of roots mentioned in the foregoing paragraphs). 
One of a doublet corresponds to an intersection of a sharply falling part of the self-energy with the $\omega$ line and thus has a positive infinitesimal residue,
while the other intersection occurs at an almost vertically {\it rising} part of the self-energy, whose residue is negative and thus nonphysical. 
These intersections almost coincide with bare 2p1h or 2h1p HF orbital energy differences, thus accounting for little to no correlation effects.

\begin{figure}
  \includegraphics[scale=0.65]{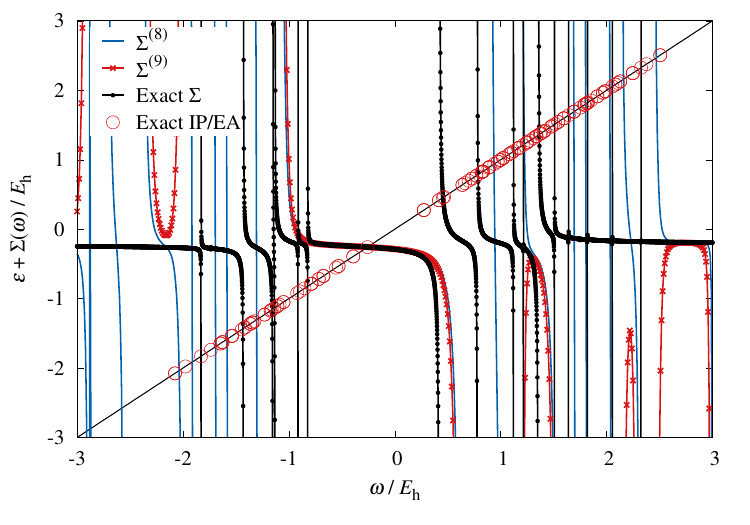}
\caption{The third diagonal elements of the exact $\bm{\epsilon}+\bm{\Sigma}(\omega)$ and $\bm{\epsilon}+\bm{\Sigma}^{(n)}(\omega)$ ($n = 8, 9$) 
 as functions of  $\omega$ for the BH molecule. The exact IPs and EAs obtained by FCI are superposed.}
\label{fig:BH_HOMOself89}
\end{figure}

\begin{figure}
  \includegraphics[scale=0.65]{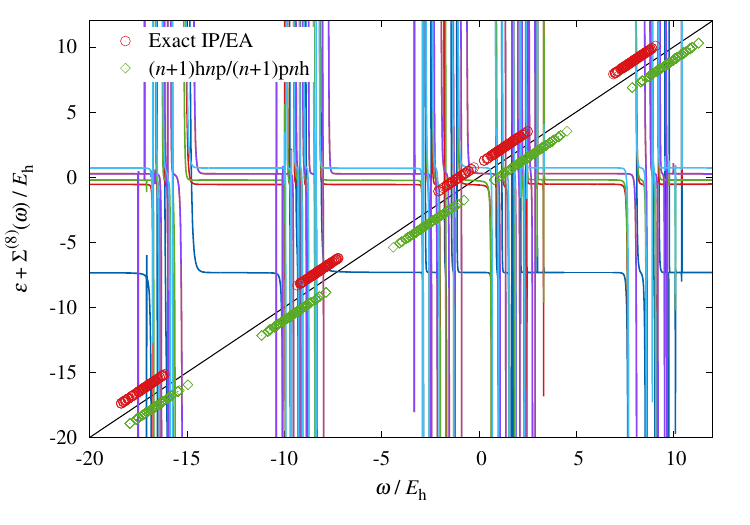}
\caption{All six diagonal elements of $\bm{\epsilon}+\bm\Sigma^{(8)}(\omega)$ 
 as functions of  $\omega$ for the BH molecule. The ($n$+1)p$n$h and ($n$+1)h$n$p HF orbital energy differences ($n = 1, 2, 3$) 
 as well as the exact IPs and EAs obtained by FCI are superposed. }
\label{fig:BH_ALLself8}
\end{figure}

These observations can be generalized to all orders. One can expect an even-order self-energy to have the same qualitatively correct functional form 
as the exact self-energy, while
the third and higher odd-order self-energies to display qualitatively wrong functional forms. This is borne out in Fig.\ \ref{fig:BH_HOMOself89}, 
in which the eighth- and ninth-order self-energies
are plotted along with the ($n$+1)h$n$p and ($n$+1)p$n$h HF orbital energy differences.  In Fig.\ \ref{fig:BH_ALLself8}, all six diagonal elements of $\bm{\Sigma}^{(8)}$  are drawn in a full domain of $\omega$.

In Fig.\ \ref{fig:BH_HOMOself89}, the odd-order $\bm{\Sigma}^{(9)}$ has a concave or convex shape in each bracket separated by its singularities and, as a result, 
 intersects the $\omega$ line only once in the shown $\omega$ domain.  In the same figure, the exact self-energy and exact IPs and EAs are overlaid, indicating
 that there should be far more intersections than just one in this $\omega$ domain. The missing satellite roots of MBGF(9) are believed to be complex.

The eighth-order self-energy is less problematic in this regard, but it comes with new issues. Its brackets are separated by the
2p1h, 2h1p, 3p2h, 3h2p, 4p3h, 4h3p, 5p4h, and 5h4p HF orbital energy differences. The even-order $\bm{\Sigma}^{(8)}$ is technically monotonically decreasing within 
each bracket, but it decreases so sharply in many brackets that it consists of near vertical lines, 
where it intersects the $\omega$ line, as seen in Figs.\ \ref{fig:BH_HOMOself89} and \ref{fig:BH_ALLself8}.
Therefore, although MBGF(8) has real roots, most of them are phantom poles with zero residues, occurring somewhere near the midpoint 
of each bracket.

Consequently, the distribution of the roots of the eighth-order inverse Dyson equation (the intersections
of $\bm{\epsilon} + \bm{\Sigma}^{(8)}$ with the $\omega$ line in Fig.\ \ref{fig:BH_ALLself8}) matches more closely 
the mere HF orbital energy differences (the green diamond plots in the same figure) than the exact roots (the red circles). 
Even though these MBGF(8) roots are not equal to the HF orbital energy differences, they are confined within brackets demarcated by these differences, which
form dense manifolds, and tend to agree with them. These satellite roots therefore include little to no electron-correlation effects, 
not to mention that most of them hardly exist physically because their residues are zero. 

These are disheartening results considering that the eighth- or ninth-order perturbation approximations 
are usually numerically exact in most other physics cases \cite{laidig,hirata_cc}. 

Our numerical tests for other molecules (not shown) have indicated that  an odd-order self-energy can generally be concave, convex, or even monotonically {\it increasing} 
within a bracket. 
Therefore, an odd-order inverse Dyson equation can have
real roots (sometimes more than one in a bracket), but the corresponding residues may fall outside the valid range of zero to one.

\begin{figure}
  \includegraphics[scale=0.65]{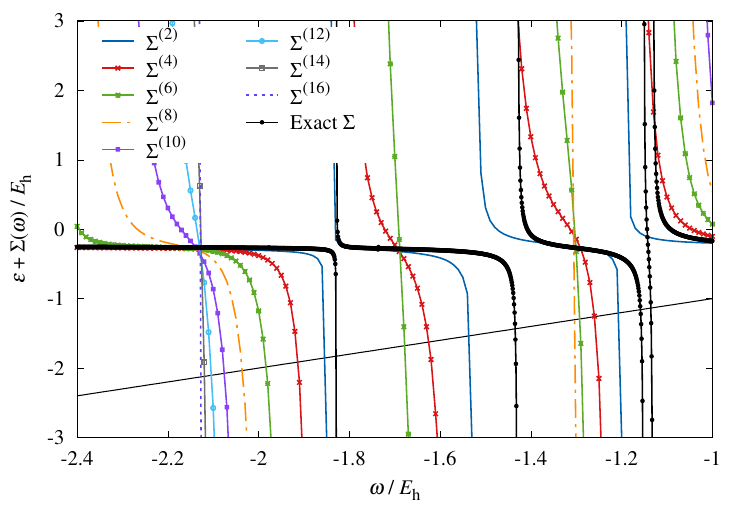}
\caption{The third diagonal elements of the exact $\bm{\epsilon}+\bm{\Sigma}(\omega)$ and $\bm{\epsilon}+\bm{\Sigma}^{(n)}(\omega)$ ($n = 2, 4, 6, \dots, 16$) 
 as functions of  $\omega$ for the BH molecule. 
 }
\label{fig:BH_HOMO2-16}
\end{figure}

One might still wonder if selectively raising only the even perturbation order may avoid these pathologies. 
Figure \ref{fig:BH_HOMO2-16} shows the evolution of $\bm{\Sigma}^{(2n)}$ as a function of the even perturbation order $2n$. 
With increasing order, the self-energy becomes more and more vertical and less shoulder-like with the intersections with the $\omega$ line
moving away from the correct locations. The corresponding residues will approach zero. 
Ironically, it appears as though the 
perturbative self-energies converged at the exact one as the perturbation order is lowered! This figure is the most compelling piece of evidence demonstrating 
the nonconvergence of the Feynman--Dyson perturbation expansion
of the self-energy. 

\begin{figure}
  \includegraphics[scale=0.65]{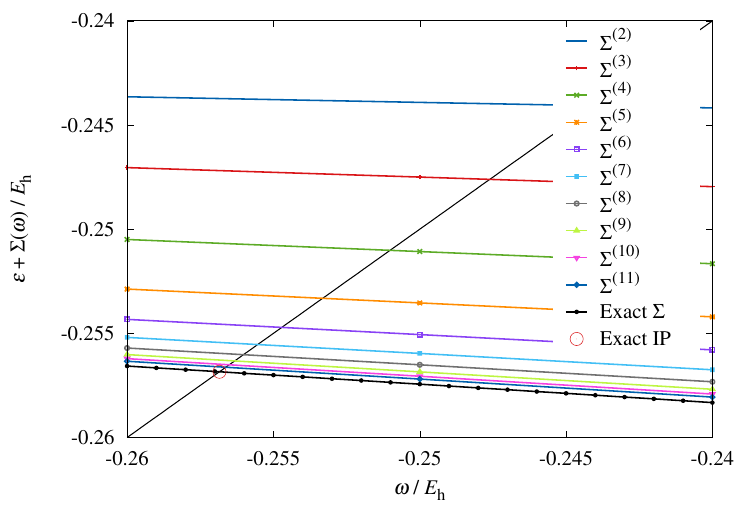}
\caption{The third eigenvalues of the exact $\bm{\epsilon}+\bm{\Sigma}(\omega)$ and $\bm{\epsilon}+\bm{\Sigma}^{(n)}(\omega)$ ($n = 2, 3, 4, \dots, 11$) as functions of  $\omega$ for the BH molecule.
The corresponding exact IP obtained from FCI is superposed.}
\label{fig:BH_HOMOself2-9}
\end{figure}

This conclusion does not undermine the utility or rapid convergence of MBGF($n$) for roots falling well within the central bracket bounded by
the highest 2h1p and lowest 2p1h energies as well as in some other $\omega$ domains (see Sec.\ \ref{sec:discussion}). 
This bracket typically encloses most principal IPs and EAs (but not core IPs) as well as low-energy satellite poles. 
Figure \ref{fig:BH_HOMOself2-9} is a close-up of the self-energy-versus-$\omega$ plot for MBGF($n$) ($2 \leq n \leq 11$) as well as the exact MBGF. 
It shows that the intersections of the self-energy and $\omega$ line systematically approach the exact IP for the HOMO.

\section{Infinite partial resummations of propagator diagrams\label{sec:infinite}}

\subsection{Vertex renormalization}

To avoid malaise of truncated perturbation approximations, infinite partial resummations of diagrams have often been invoked. One way of doing 
this is by renormalizing vertexes. 

Here, a ``vertex'' refers to 
an instantaneous Coulomb interaction denoted by a horizontal (dashed) line in a Hugenholtz--Shavitt--Bartlett diagram  
and is equal to an anti-symmetrized two-electron integral containing 
both the direct {\it and} exchange contributions \cite{szabo,shavitt}. It should be distinguished from a spacetime point of contact between propagator lines and an interaction line
in a Goldstone diagram, each standing for only the direct {\it or} exchange contribution, often used in the $GW$ approximations \cite{Hedin,GWLouie,GW1,GW2,Vlcek2019,Vlcek2022}.

\begin{figure}
  \includegraphics[scale=0.3]{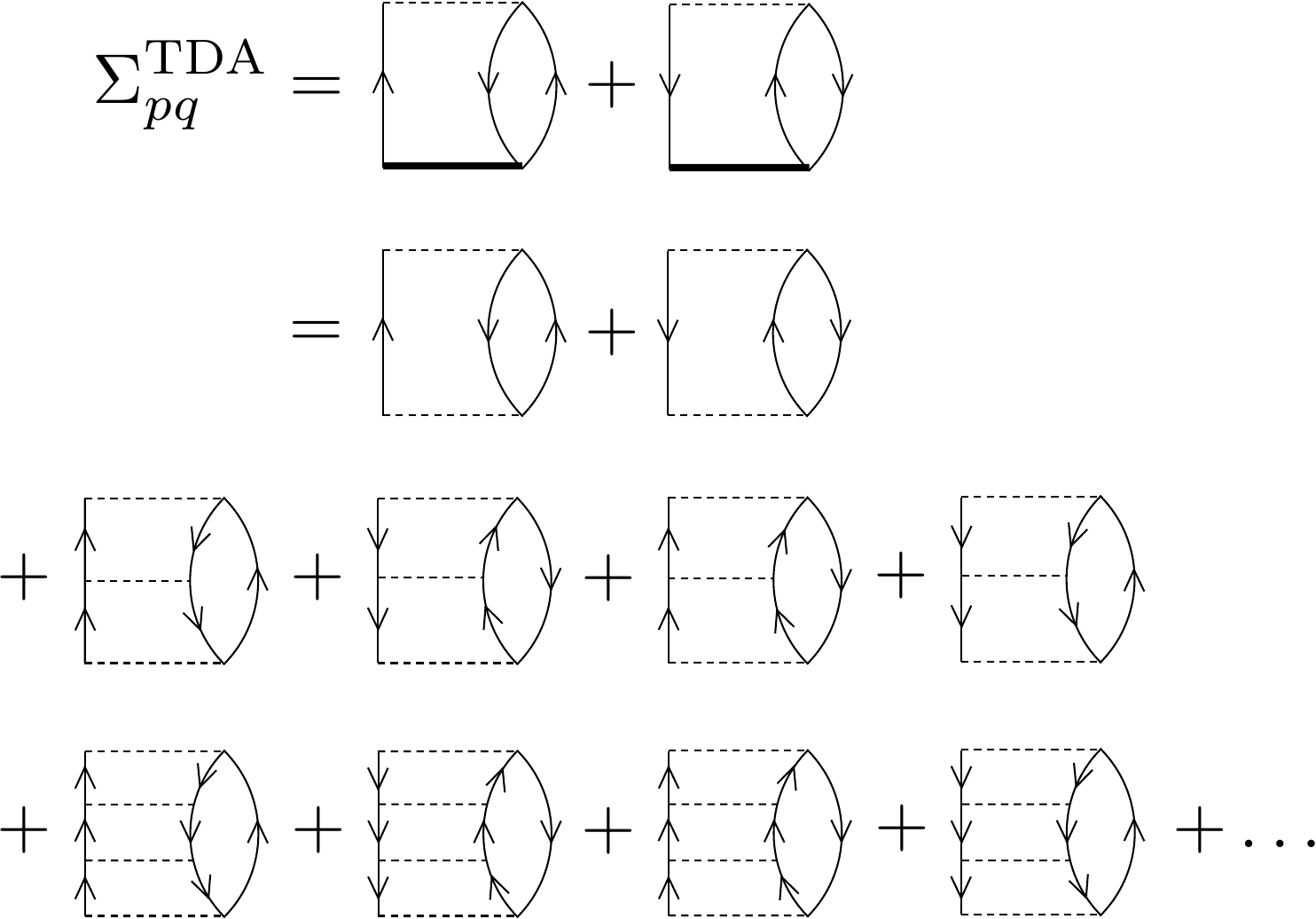}
\caption{The self-energy in the TDA approximation.}
\label{fig:TDA2}
\end{figure}

\begin{figure}
  \includegraphics[scale=0.3]{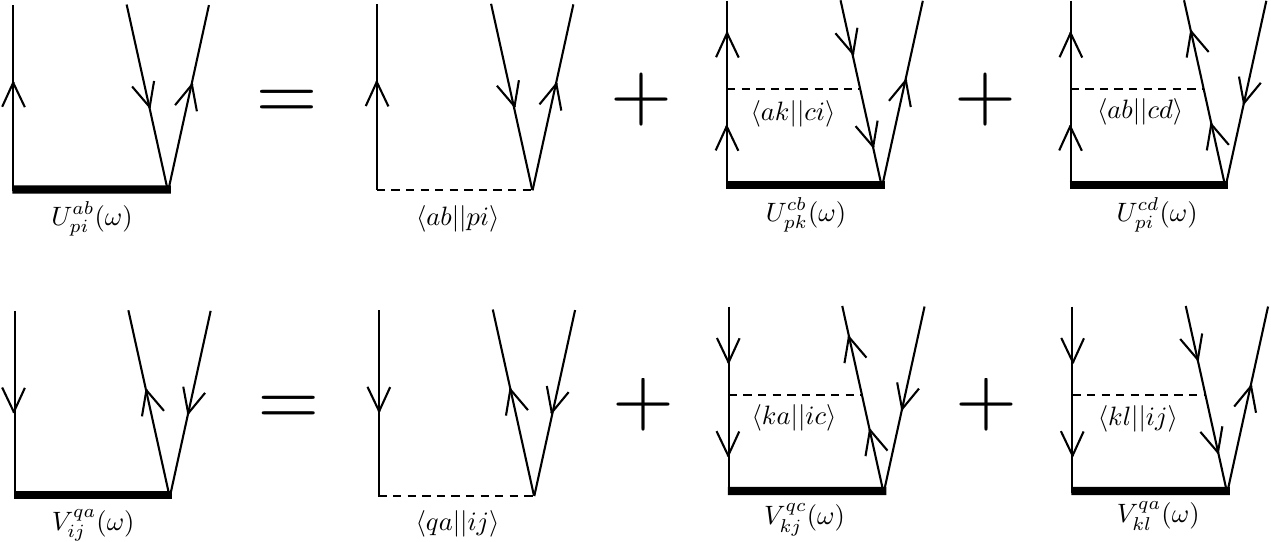}
\caption{The diagrammatic equations for the bold-line (renormalized) vertexes of TDA.}
\label{fig:TDA2vertex}
\end{figure}

In the two-particle-hole Tamm--Dancoff approximation (TDA) \cite{Linderberg67,PurvisOhrn1974,Schirmer1978,Walter1981}, also known as 
the Brueckner--Hartree--Fock method \cite{Day1967,Baldo_book} or the T approximation \cite{Kadanoff_book}, the 
diagrams of the types in Fig.\ \ref{fig:TDA2} are summed over up to an infinite order. In this figure, 
the first two diagrams in the third and fourth lines are ``ring'' diagrams, and 
the subsequent two diagrams are ``ladder'' diagrams. 
The effect of this infinite resummation 
can be folded into the bold-line vertexes as appearing in the first line of Fig.\ \ref{fig:TDA2}, which must, in turn, satisfy the diagrammatic equations drawn in Fig.\ \ref{fig:TDA2vertex}. 
They take the form of the amplitude equations of  coupled-cluster theory \cite{bartlettRMP,shavitt} and are written algebraically as
\begin{eqnarray}
\Big(\omega+\epsilon_i - \epsilon_a - \epsilon_b\Big) U^{ab}_{pi}(\omega) &=& \langle ab || pi \rangle -  P(ab) \sum_{c,k} \langle ak || ci \rangle U^{cb}_{pk}(\omega) \nonumber\\&&+ \frac{1}{2} \sum_{c,d} \langle ab || cd \rangle U^{cd}_{pi}(\omega) , \label{eq:TDA1_1} \\
\Big(\epsilon_i + \epsilon_j - \omega - \epsilon_a\Big) V^{qa}_{ij}(\omega) &=& \langle qa || ij \rangle - P(ij) \sum_{c,k} \langle ka || ic \rangle V^{qc}_{kj}(\omega) \nonumber\\&&+ \frac{1}{2} \sum_{k,l} \langle kl || ij \rangle V^{qa}_{kl}(\omega) , \label{eq:TDA1_2}
\end{eqnarray}
where $P(ab)$ is an antisymmetrizer \cite{shavitt}. Unlike coupled-cluster theory, where there is only one type of cluster excitation amplitudes (denoted by $T$), there
are two (2p1h and 2h1p) types of modified vertexes whose numerical values are stored in $\bm{U}$ and $\bm{V}$. 
They represent electron-electron repulsion tempered by 
screening and other higher-order electron-correlation effects. 

Equations (\ref{eq:TDA1_1}) and (\ref{eq:TDA1_2}) are a system of linear equations, which can, therefore, be ill-conditioned at some $\omega$'s. 
In this work, they are solved in an iterative algorithm; i.e., starting with initial guesses 
of $\bm{U}$ and $\bm{V}$, we substitute them in the right-hand sides of these equations to update $\bm{U}$ and $\bm{V}$ in the left-hand sides, and repeat this process 
until convergence. Therefore, in practice, the highest order of the ring and ladder diagrams that are actually included in the calculation 
is capped by the number of cycles taken in this iterative solution. 
Upon convergence, the self-energy is obtained as
\begin{eqnarray}
\Sigma^{\text{TDA}}_{pq}(\omega) = \frac{1}{2} \sum_{i,a,b} \langle qi || ab \rangle U^{ab}_{pi}(\omega) - \frac{1}{2} \sum_{i,j,a} \langle ij || pa \rangle V^{qa}_{ij}(\omega). 
\nonumber\\ \label{eq:Sigma_TDA}
\end{eqnarray}
See Ref.\ \cite{Ortiz2023} for a more efficient and stable algorithm, which, however, is not exempt
from the pathological behaviors discussed below, which are deeply rooted in the formalism.

For a total energy, an infinite partial resummation of the corresponding ring and ladder diagrams defines an instance of coupled-cluster theory \cite{bartlettRMP,shavitt} known as 
linearized coupled-cluster doubles (LCCD) or by other names \cite{LCPMET,CPA0,Bartlett1977,Bartlett1978,Bartlett1979,Ahlrichs1979,Koch1981}. This method is $\omega$-independent and is thus free from the nonconvergence problems
discussed here. It instead displays other problems concerning accuracy \cite{Taube2009}.

\begin{figure}
  \includegraphics[scale=0.65]{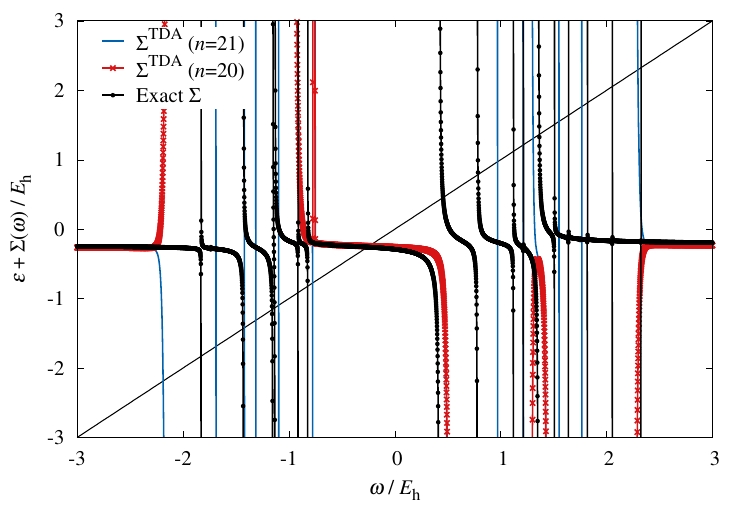}
\caption{The exact ${\epsilon_3}+{\Sigma_{33}}(\omega)$ and ${\epsilon_3}+\Sigma_{33}^{\text{TDA}}(\omega)$ as functions of  $\omega$ for the BH molecule. The self-energy in TDA is determined after $n$ cycles of the iterative solution of the amplitude equation, 
summing over the ring and ladder diagrams through the order $n+1$.}
\label{fig:BH_HOMOselfTDA}
\end{figure}

Figure \ref{fig:BH_HOMOselfTDA} compares the self-energy of  TDA  with the exact self-energy. The former 
was obtained after either 20 or 21 cycles of the iterative
solution of the amplitude equations, summing over ring and ladder diagrams through the order 21 or 22, respectively.  As may be expected from the discussion on perturbative self-energies,  TDA 
does not improve the overall appearance of the self-energy; rather, it  degrades it.
 
For instance, $\bm{\Sigma}^{\text{TDA}}$ obtained after 20 cycles has qualitatively wrong (concave and convex) functional forms (outside the central bracket)
because computationally it is a high-odd-order perturbation theory.  In contrast, $\bm{\Sigma}^{\text{TDA}}$ after 21 cycles has qualitatively correct functional forms (as it is
a high-even-order perturbation theory),
but it consists of numerous vertical lines. 
Therefore, TDA predicts vastly different IPs and EAs depending on 
the number of cycles taken in the iterative solution---an artifact of calculations---and is therefore methodologically ill-defined insofar as all roots are sought. 
Furthermore, both of these conflicting predictions are equally meaningless. After 20 cycles, there are no real roots outside the central bracket in this graph; after 21 cycles, 
most roots are phantom poles with zero residues. 
However, this observation should not be misconstrued to mean that TDA is useless; for the principal IPs and EAs, TDA is sound and useful. 

It is unsurprising for an infinite partial resummation to inherit the pathologies of higher-order perturbation theory, but
this does not imply that the same pathologies plague all vertex renomalization techniques. The equation-of-motion (EOM) formalism of Green's function theory \cite{Rowe1968,simons_3rd,herman,Herman1980,Herman1980_2} 
also involves an infinite partial resummation of diagrams, but it can be recast into a matrix diagonalization \cite{BakerPickup} and  may thus be free from the type of the nonconvergence problems discussed above. 
The random-phase approximation (RPA) \cite{Dickhoff_chapter7}, which is a lowest-order member of the EOM approximation series, as well as the related $GW$ 
approximations \cite{Hedin,GWLouie,GW1,GW2,Vlcek2019,Vlcek2022} and  the
frequency-dependent part of the ADC self-energy \cite{schirmer1982,schirmer} may also be more robust than TDA, 
although this has not been numerically confirmed nor is the convergence toward exactness implied.
The Parquet method \cite{Bergli2011} sums over a related, but distinct class of diagrams and hence the present conclusion may not  apply, either.

The EOM-CC method \cite{stanton_eom,Stanton_ip,Nooijen_ea2,Bartlett_ip,Stanton_ip2,Gour2005,Kamiya_ip,Gour2006,Kamiya_ea,hirata_ipeomcc,Bartlett2012} has been argued  to be 
a coupled-cluster Green's function \cite{Gunnarsson1978,nooijen_gf1,nooijen_gf2,meissner1993,Kowalski1,Kowalski2,Kowalski3,Peng2019,Peng2020,Rehr2020,Pathak2023} and can also be viewed as an infinite resummation of the propagator diagrams \cite{deltamp}. 
However, this method does not suffer from the pathologies discussed above; it instead yields IPs and EAs for all (principal, satellite, and even core-ionized \cite{Vidal2019}) 
states that are systematically convergent 
at the exact values with increasing excitation level. 
EOM-CC separates the IP and EA sectors and its roots are obtained as eigenvalues of a non-Hermitian matrix. 
Like MBGF($n$), the EOM-CC roots are not guaranteed to be real, but they are almost never complex in practice.

It may be inferred that the EOM formalisms of Green's functions, be they based on perturbation theory or coupled-cluster theory, are more robust 
and potentially converging because their IPs and EAs are obtained directly as eigenvalues of some matrix rather than by solving an 
inverse Dyson equation. In other words, the EOM formalisms may attain stability and convergence 
by divorcing from Green's function theory in the Feynman--Dyson style. 
That the exact MBGF and MBGF(2), the latter being the only Feynman--Dyson perturbation order largely exempt from the pathologies, 
can also be recast into a matrix eigenvalue problem \cite{Walter_JMathPhys}
is consistent with this inference.

\subsection{Edge renormalization} 

\begin{figure}
  \includegraphics[scale=0.3]{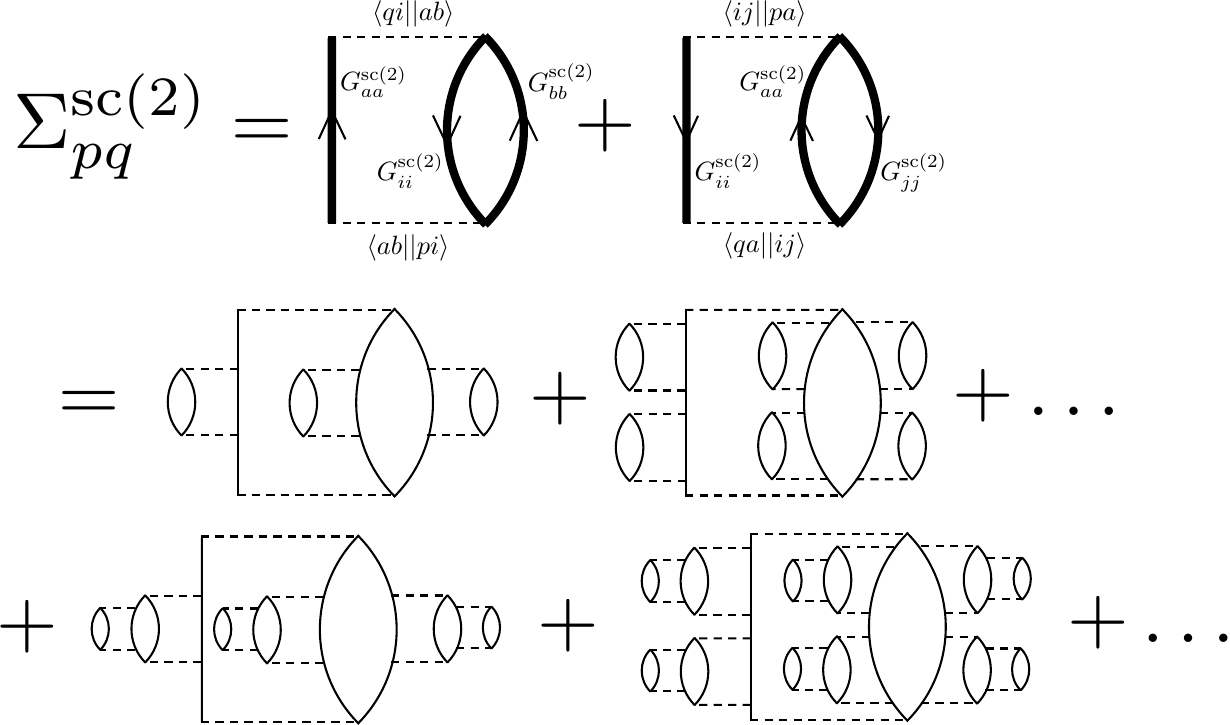}
\caption{The self-consistent second-order self-energy.}
\label{fig:SelfconsistentSigma2}
\end{figure}

\begin{figure}
  \includegraphics[scale=0.3]{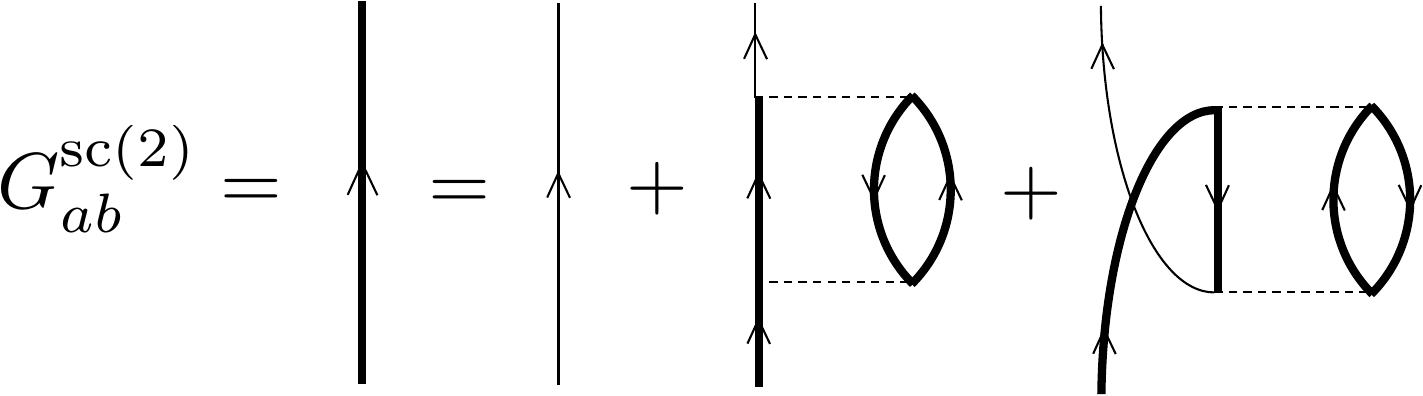}
\caption{The diagrammatic equation for the bold-line (renormalized) edge of the self-consistent second-order Green's function method.}
\label{fig:SelfconsistentG2}
\end{figure}

Another way of performing a diagram resummation is by renormalizing edges. 
By replacing all three edges in each diagram of $\delta\bm{\Sigma}^{(2)}$ by the corresponding bold-line Green's functions $\bm{G}^{\text{Dyson}(2)}$ 
of Fig.\ \ref{fig:GDyson2}, we include an infinite number of ``row-house'' diagrams appearing in the second line of Fig.\ \ref{fig:SelfconsistentSigma2}. 
Furthermore, if the bold-line Green's function (designated by $\bm{G}^{\text{sc}(2)}$) 
comes from this very edge-modified self-energy $\bm{\Sigma}^{\text{sc}(2)}$, 
as in Fig.\ \ref{fig:SelfconsistentG2}, we account for another infinite set of ``tower'' diagrams shown in the third line of  Fig.\ \ref{fig:SelfconsistentSigma2}. 
This self-consistency between the self-energy and Green's function was emphasized by  
Baym and Kadanoff \cite{BaymKadanoff1961,Baym_selfconsistent,Kadanoff_book,Dahlen2005} as an essential ingredient of an approximate MBGF method that obeys conservation laws.
(The first-order self-consistent MBGF method is identified as the HF theory \cite{Dickhoff_chapter7}.)
In chemistry, approximations inspired by this idea are enjoying a revival for its possible ability to describe strong correlation more accurately \cite{Surjan1998,Holleboom1990,Phillips2014,CoveneyTew2023}.

In the diagonal approximation, a {\it self-consistent} second-order self-energy \cite{VanNeck1991} is thus defined by the same equation as Eq.\ (\ref{eq:contour1}) but with each $\bm{G}^{(0)}$ replaced by 
$\bm{G}^{\text{sc}(2)}$,
\begin{widetext}
\begin{eqnarray}
\Sigma_{pq}^{\text{sc}(2)}(\omega) 
&=& (-1)^1 i \sum_i^{\text{occ.}} \sum_{a<b}^{\text{vir.}} 
\int_{-\infty}^{\infty}\frac{d\omega_a}{2\pi} 
\int_{-\infty}^{\infty}\frac{d\omega_i}{2\pi} 
\, {(-i)\langle qi||ab \rangle}{(-i)\langle ab|| pi \rangle}   
\,  iG_{aa}^{\text{sc}(2)}(\omega_a) iG_{ii}^{\text{sc}(2)}(\omega_i) iG_{bb}^{\text{sc}(2)}(\omega+\omega_i-\omega_a) \nonumber\\
&& + (-1)^1 i  \sum_a^{\text{vir.}} \sum_{i<j}^{\text{occ.}} 
\int_{-\infty}^{\infty}\frac{d\omega_i}{2\pi} 
\int_{-\infty}^{\infty}\frac{d\omega_a}{2\pi} 
\, (-i) \langle qa|| ij \rangle (-i) \langle ij||pa \rangle 
\,  iG_{ii}^{\text{sc}(2)}(\omega_i) iG_{aa}^{\text{sc}(2)}(\omega_a) iG_{jj}^{\text{sc}(2)}(\omega+\omega_a-\omega_i) \label{eq:contour2}\\
&=& \frac{1}{2} \sum_i^{\text{occ.}} \sum_{a,b}^{\text{vir.}}  \sum_{I(i)}^{\text{IP}} \sum_{A(a),B(b)}^{\text{EA}} \frac{\langle qi||ab \rangle\langle ab|| pi \rangle  }{\omega + \omega_{I(i)}-\omega_{A(a)}-\omega_{B(b)}} F(\omega_{I(i)}) F(\omega_{A(a)}) F(\omega_{B(b)})
\nonumber\\&& 
+ \frac{1}{2} \sum_a^{\text{vir.}} \sum_{i,j}^{\text{occ.}} \sum_{I(i),J(j)}^{\text{IP}} \sum_{A(a)}^{\text{EA}} \frac{\langle qa|| ij \rangle \langle ij||pa \rangle }{\omega + \omega_{A(a)}-\omega_{I(i)}-\omega_{J(j)}}
F(\omega_{A(a)}) F(\omega_{I(i)}) F(\omega_{J(j)}), 
\label{eq:sc2}
\end{eqnarray}
\end{widetext}
where the occupied spinorbital index $i$ and the imaginary unit $i$ are to be distinguished, and 
$\omega_{I(q)} (< 0)$ and $\omega_{A(q)} (> 0)$ are an IP and EA  root, respectively, of the inverse Dyson equation in the diagonal approximation, which satisfy
\begin{eqnarray}
\epsilon_q + \Sigma_{qq}^{\text{sc}(2)}(\omega_{I(q)}) &=& \omega_{I(q)} ,  \\
\epsilon_q + \Sigma_{qq}^{\text{sc}(2)}(\omega_{A(q)}) &=& \omega_{A(q)} .
\end{eqnarray}
This ansatz may differ somewhat from those of Van Neck {\it et al.}\ \cite{VanNeck1991} or of Dahlen and van Leeuwen \cite{Dahlen2005} as it involves 
some additional approximations, but these differences have no impact on the following analysis.
The corresponding residues are given by
\begin{eqnarray}
F(\omega_{I(q)}) \equiv \text{Res}_{\omega_{I(q)}}  G_{qq}^{\text{sc}(2)}(\omega) = \left\{ 1 - \left. \frac{\partial \Sigma_{qq}^{\text{sc}(2)}(\omega)}{\partial \omega}\right|_{\omega_{I(q)}} \right\}^{-1}.
\nonumber\\ \label{eq:sc2residue}
\end{eqnarray}
These conditions imply the self-consistency,
\begin{eqnarray}
\bm{G}^{\text{sc}(2)}(\omega) &=& \left\{ \omega\bm{1} - \bm{\epsilon} - \bm{\Sigma}^{\text{sc}(2)}(\omega) \right\}^{-1}, 
\end{eqnarray}
in the diagonal approximation, although an explicit evaluation of $\bm{G}^{\text{sc}(2)}$ is never needed.

In practice, Eqs.\ (\ref{eq:sc2})--(\ref{eq:sc2residue}) are solved iteratively. In cycle  zero ($n=0$), we use $\bm{G}^{(0)}$ in the right-hand side 
of Eq.\ (\ref{eq:sc2}) and obtain $\bm{\Sigma}^{\text{sc}(2)} (n=0) = \bm{\Sigma}^{(2)}$ in the left-hand side.
In cycle one, therefore, the new Green's function is $\bm{G}^{\text{Dyson}(2)}$. By determining all of its poles and residues and substituting them 
back into Eq.\ (\ref{eq:sc2}), we obtain $\bm{\Sigma}^{\text{sc}(2)} (n=1)$. In cycle two, similarly, we get $\bm{\Sigma}^{\text{sc}(2)} (n=2)$, and so on. In each cycle, figuratively speaking,
a ``new floor'' is added to each of the ``tower'' diagrams (see the third line of Fig.\ \ref{fig:SelfconsistentSigma2}).
This calculation quickly becomes intractable because the number of poles increases
factorially with iterative cycles. Dahlen and van Leeuwen \cite{Dahlen2005} and Zgid and coworkers \cite{Phillips2014,Kananenka2016,Lan2016,Neuhauser2017} devised
 imaginary-time-dependent (or finite-temperature) algorithms for this method, which seem to overcome this  computational intractability.

\begin{figure}
  \includegraphics[scale=0.65]{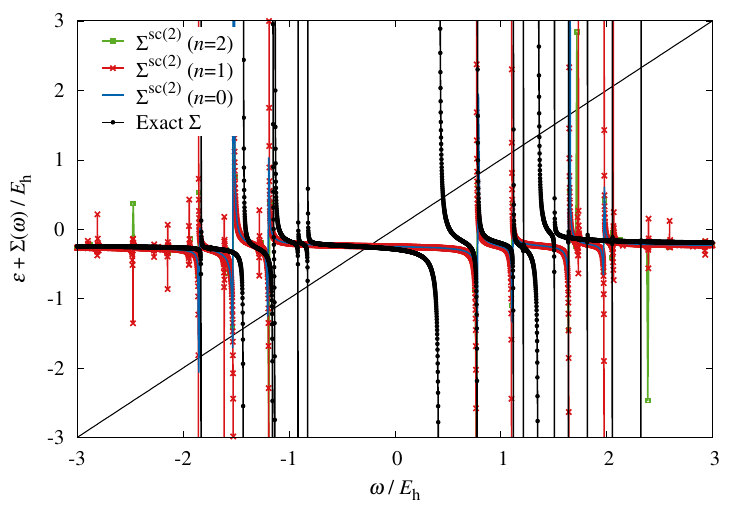}
\caption{The exact ${\epsilon_3}+{\Sigma_{33}}(\omega)$ and ${\epsilon_3}+\Sigma_{33}^{\text{sc}(2)}(\omega)$ as a function of $\omega$ for the BH molecule. The self-consistent self-energy is determined after $n$ cycles of the iterative $i$th ($a$th) edge replacement by diagonal bold-line ${G}_{ii}^{\text{sc}(2)}$ (${G}_{aa}^{\text{sc}(2)}$). $\bm\Sigma^{\text{sc}(2)}$ ($n=0$) corresponds to the unmodified $\bm\Sigma^{(2)}(\omega)$.}
\label{fig:BH_HOMOselfSC2}
\end{figure}

Figure \ref{fig:BH_HOMOselfSC2} compares the self-energies of the self-consistent second-order Green's function method after zeroth, first, and second iterative cycles. 
In each case, the self-energy has the qualitatively correct functional form, i.e., monotonically decreasing in each bracket separated by its singularities.
This is because the self-energy expression [Eq.\ (\ref{eq:sc2})] is isomorphic to the second-order self-energy, which has only the first-order poles. 
Furthermore, unlike higher-order perturbative self-energies or TDA, the brackets' boundaries shift from one cycle to the next,
and hence the satellite roots are no longer confined by brackets of the mere HF orbital energy differences. 

Nonetheless, this does not mean that these satellite roots are improved. To the contrary, they seem to deteriorate with increasing iterative cycles.
Figure \ref{fig:BH_HOMOselfSC2} shows that in the first and second cycles, new singularities of the self-energy emerge, e.g., in the domain $-3\,E_{\text{h}} \leq \omega \leq -2\,E_{\text{h}}$, where
there are no corresponding singularities of the exact self-energy. The mechanism by which these spurious singularities 
multiply rapidly with increasing self-consistent cycles is as follows:
In each cycle, the poles of the Green's function define new $\omega$ brackets, in each of which
there is one root of the inverse Dyson equation. These brackets are demarcated by singularities of the self-energy, Eq.\ (\ref{eq:sc2}), which are the 2p1h ($\omega_{A(a)}+\omega_{B(b)}-\omega_{I(i)}$) or 2h1p ($\omega_{I(i)}+\omega_{J(j)}-\omega_{A(a)}$) energy differences of the poles of the Green's function (which are no longer the HF orbital energy differences after the first cycle). As the number of poles increases,
the number of brackets and thus the number of roots increase extremely rapidly, quickly exceeding the correct total number of roots in a finite basis set. The growth is factorial of the iterative cycle.

\begin{figure}
  \includegraphics[scale=0.65]{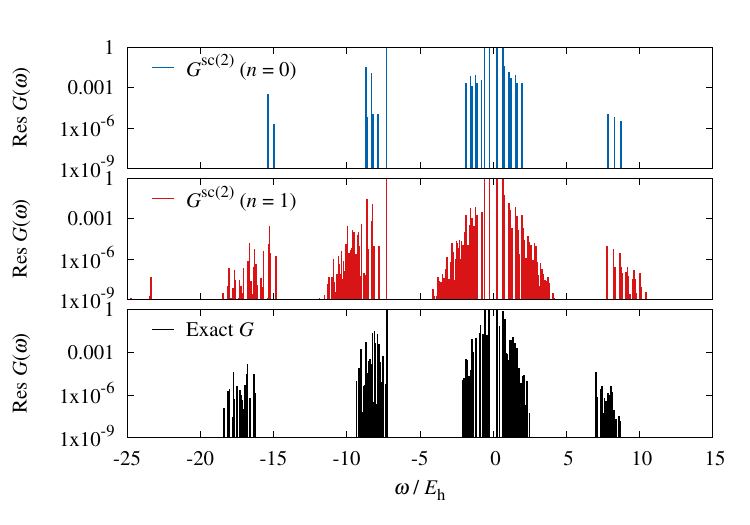}
\caption{Residues at poles $\omega$  of $\bm{G}^{\text{sc}(2)}(\omega)$ after $n$ cycles of the self-consistent iteration as well as of 
the exact $\bm{G}(\omega)$ for the BH molecule. $\bm{G}^{\text{sc}(2)}(\omega)$ at $n=0$ corresponds to $\bm{G}^{\text{Dyson}(2)}(\omega)$.}
\label{fig:BH_SCresidues}
\end{figure}

Figure \ref{fig:BH_SCresidues} shows the poles and residues in the zeroth and first cycles of the self-consistent second-order Green's function method
in comparison with the exact poles and residues.  The bottom panel is the same as Fig.\ \ref{fig:BH_residues}, but shown in a full  domain of $\omega$,
and the top panel is equivalent to MBGF(2) in the diagonal approximation. In the zeroth cycle ($n=0$), the distribution of the poles agree reasonably well
with that of the exact poles. The number of poles is 72 as compared with 600 exact poles including ones with zero residues. In the first cycle ($n=1$), the number of 
poles already reaches 4,314, far exceeding the total number (600) of ionized and electron-attached states of a FCI calculation. Furthermore, these spurious
poles are not necessarily phantom mathematical roots with zero residues; they have nonzero residues and
encroach on the regions where there are no exact poles. In the second cycle ($n=2$), the number of roots reaches such an astronomical value that our computer code can no longer handle, and we
judged that it was not worthwhile to pursue full self-consistency. This difficulty was recognized by earlier workers, many of whom then decided to abandon this class of methods.

The bold-line self-energy diagrams in Fig.\ \ref{fig:SelfconsistentSigma2} is obtained by cutting one bold line (and then trimming the resulting two dangling lines) \cite{Hirata2017} 
of a skeleton diagram of the Luttinger--Ward functional \cite{luttingerward,BaymKadanoff1961,Baym_selfconsistent,Lin2018} in the zero-temperature limit. 
Several mathematical difficulties associated with this functional have been 
reported recently \cite{Schafer2013,Stan2015,Kozik2015,Rossi2015,Schafer2016,Tarantino2017,Gunnarsson2017}.

\section{Analysis\label{sec:analysis}}

\subsection{Cause of nonconvergence\label{sec:discussion}}
 
What is the cause of the nonconvergence? This can be answered by analyzing a model Green's function of the form \cite{hiratapra},
\begin{eqnarray}
g(\omega) &=& \frac{1}{\omega-E_1}+\frac{1}{\omega-E_2}+\frac{1}{\omega-E_3}+\frac{1}{\omega-E_4}, \label{eq:g}
\end{eqnarray}
which consists of four poles. These poles are stipulated to be quadratic functions of the perturbation strength $\lambda$.
\begin{eqnarray}
E_1 &=& 1.9 + 0.2\lambda + 0.2 \lambda^2,\\
E_2 &=& 0.75 + 0.1\lambda + 0.1 \lambda^2,\\
E_3 &=& -1.1 - 0.1\lambda - 0.1 \lambda^2,\\
E_4 &=& -2.2 - 0.15\lambda - 0.15 \lambda^2.
\end{eqnarray}
The functional forms of $g(\omega)$ and $E$'s are arbitrary, but a different choice would lead to essentially the same conclusion. 

One can expand $g(\omega)$ in a Taylor series in $\lambda$, 
\begin{eqnarray}
g(\omega) &=& g^{(0)}(\omega) + \lambda g^{(1)}(\omega) + \frac{\lambda^2}{2!} g^{(2)}(\omega) + \frac{\lambda^3}{3!} g^{(3)}(\omega) + \dots. \nonumber\\
\end{eqnarray}
A truncation of this series after a finite number of terms captures the key features of the Feynman--Dyson perturbation approximations of a Green's function or self-energy. 

\begin{figure}
  \includegraphics[scale=0.65]{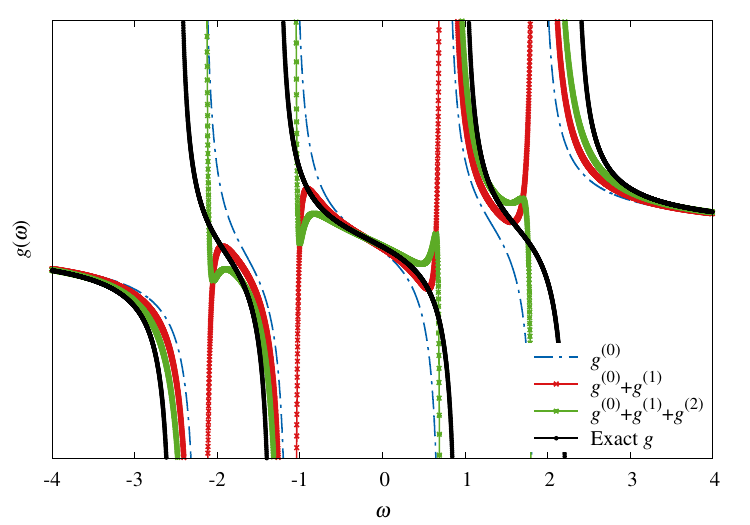}
\caption{Taylor expansions up to the second order in $\lambda$ of the model Green's function $g(\omega)$ of Eq.\ (\ref{eq:g}).}
\label{fig:Taylor1}
\end{figure}

Figure \ref{fig:Taylor1} shows the zeroth-, first-, and second-order Taylor expansions of $g(\omega)$. 
The exact $g(\omega)$ and its zeroth-order approximation $g^{(0)}(\omega)$ have essentially the same functional forms in that they are
separated into consecutive $\omega$ brackets by their singularities and within each bracket they are monotonically decreasing functions. 
The first-order approximation $g^{(0)}+g^{(1)}$ has a qualitatively different functional form, which is convex or concave except in the central bracket.
The $g^{(0)}+g^{(1)}+g^{(2)}$ largely restores the same functional form as the exact $g(\omega)$, although the former is a rather poor approximation at many frequencies despite the fact that 
the second-order Taylor expansions of $E_1$ through $E_4$ are exact. 
These are consistent with the overall patterns of behaviors of {\it ab initio} odd- and even-order perturbative self-energies observed in the foregoing sections.

\begin{figure}
  \includegraphics[scale=0.65]{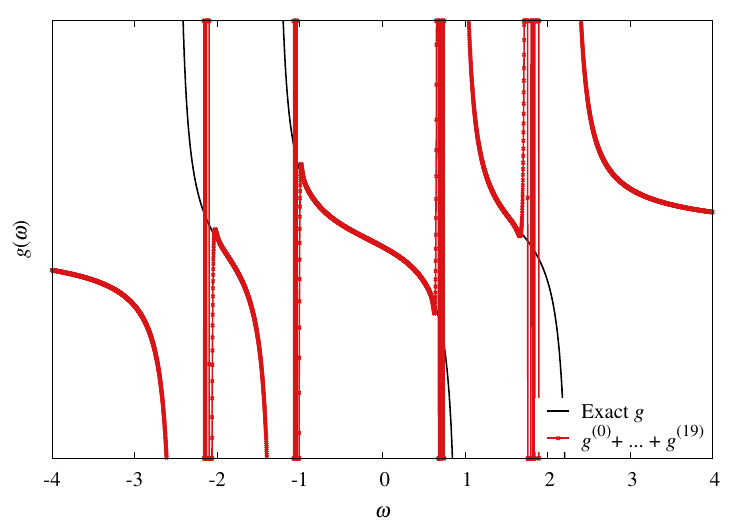}
\caption{Same as Fig.\ \ref{fig:Taylor1} but for the nineteenth-order Taylor expansion.}
\label{fig:Taylor2}
\end{figure}

Figure \ref{fig:Taylor2} extends this analysis to the nineteenth-order Taylor expansion. 
It is convergent at the exact $g(\omega)$ in some domains of $\omega$, but nonconvergent in the other domains.
Generally, convergence is attained within the overlap of the bracket demarcated by
the singularities of $g(\omega)$ and the one delineated by the singularities of $g^{(0)}(\omega)$. (The latter singularities are where
the near-vertical rapid oscillations of the nineteenth-order Taylor expansion occur.) In this overlap, both $g(\omega)$ 
and $g^{(0)}(\omega)$ are in the middle ``shoulder'' part of the monotonically decreasing functions of $\omega$, displaying similar functional forms. 
Here, a Taylor expansion from $g^{(0)}(\omega)$ is convergent. 
Outside this overlap, either $g(\omega)$ or $g^{(0)}(\omega)$ is 
in the left ``neck'' part of the function, while the other is in the right ``arm'' part. 
They are both monotonically decreasing, but one of them has a singularity at one boundary of the domain, while the other has a singularity at the other boundary, and they have dissimilar functional forms
in this sense. 
The $g(\omega)$ in this domain is nonanalytic and its Taylor expansion has zero radius of convergence. 

In a molecular Green's function or self-energy, there are dense manifolds of singularities outside the central overlapping bracket, which encloses most principal roots and some low-energy satellite roots.
Therefore, in practice, their perturbation expansions are safely convergent only in this central overlapping bracket as well as in the two terminal overlapping brackets.
The central overlapping bracket is an overlap of the bracket bounded by the highest 2p1h and 
lowest 2h1p HF orbital energy differences and the bracket demarcated typically by the least negative and least positive singularities of the exact self-energy \cite{SunBartlett1996}. Only the roots in this central overlapping bracket, be they principal or satellite, are reliably determined by MBGF($n$). 

\subsection{Pad\'{e} approximants\label{sec:pade}}

In Sec.\ \ref{sec:discussion}, the nonconvergence is shown to be caused by the very definition of the exact Green's function, which is nonanalytic. 
Even when the positions of the poles, $E_1, \dots, E_4$, are exactly expanded by second-order Taylor series, the Taylor expansion of $g(\omega)$ is 
not only inexact at the second order, but also nonconvergent at exactness at an infinite order.  Therefore, the cause of nonconvergence 
is a mathematical one:\ The rational-function form of the exact Green's function is unamenable to a converging Taylor expansion, 
even though it contains full physical information about the poles.

For this mathematical problem, the purely mathematical resummation technique of Pad\'{e} approximants \cite{Bender} may prove useful. 
The $[M,N]$ Pad\'{e} approximant of a function $f(\lambda)$ is a power-series expansion of a rational function and is defined by
\begin{eqnarray}
[M, N]_f(\lambda) = \frac{a_0 + a_1 \lambda + a_2 \lambda^2 + \dots + a_M \lambda^M}{1 + b_1 \lambda + b_2 \lambda^2 + \dots + b_N \lambda^N},
\end{eqnarray}
whose expansion coefficients, $\{a_0,\dots,a_M\}$ and $\{b_1,\dots,b_N\}$, are determined such that the above agrees with 
the $M+N+1$-order Taylor expansion of $f(\lambda)$. Pad\'{e} approximants are known to 
generate a rapidly converging series out of even a divergent Taylor series, often working   well  ``beyond their proven range of applicability'' \cite{Bender}
at essentially no additional cost if the $M+N+1$-order Taylor expansion dominates the cost.

\begin{figure}
  \includegraphics[scale=0.65]{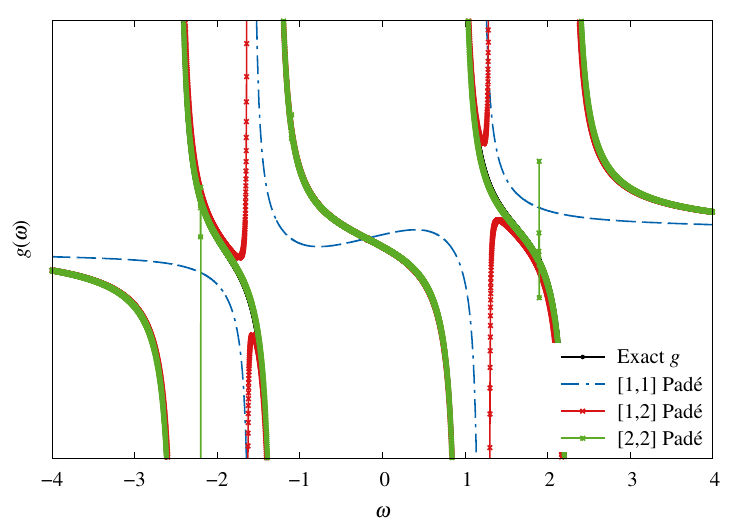}
\caption{Same as Fig.\ \ref{fig:Taylor1} but for Pad\'{e} approximants.}
\label{fig:Pade}
\end{figure}

Figure \ref{fig:Pade} shows the [1,1], [1,2], and [2,2] Pad\'{e} approximants of $g(\omega)$.
The [2,2] Pad\'{e} approximant (summing through the fifth-order Taylor series) 
is nearly exact and indistinguishable to the human eye from $g(\omega)$ except for 
a few spurious spikes. This is in contrast to Fig.\ \ref{fig:Taylor2} in which even the nineteenth-order Taylor expansion suffers from infinite errors in multiple $\omega$ domains.
The remarkable performance of Pad\'{e} approximants is ascribed to separately expanding the numerator and denominator of the rational-function form of $g(\omega)$. 

Pad\'{e} approximants were introduced to many-body perturbation energies by Goscinski \cite{Goscinski1967} and further developed by 
Br\"{a}ndas and Goscinski \cite{Brandas1970,GoscinskiBrandas1971} and by Bartlett and Br\"{a}ndas \cite{Bartlett1972,Bartlett1973}. 
They were applied to both convergent \cite{laidig} and divergent \cite{hirata_cc} series of many-body perturbation energies, and in the latter case 
a rapid convergence was restored.  In each case, the size-consistency is maintained \cite{Bartlett1972,Bartlett1973}.
Their applications to Green's functions were proposed by Goscinski and Lukman \cite{GoscinskiLukman} and  by Linderberg and Ratner \cite{Linderberg1970}, but 
no numerical results beyond a two-site model calculation were given. 

\begin{figure}
  \includegraphics[scale=0.65]{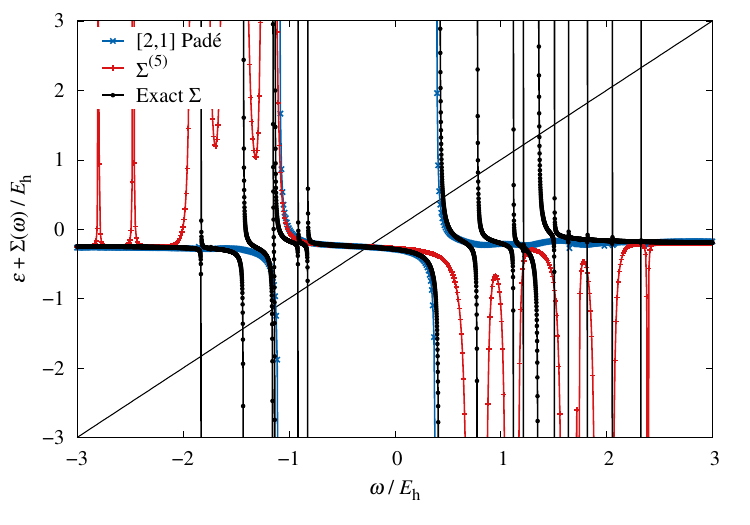}
\caption{The third diagonal element of $\bm{\epsilon} + \bm{\Sigma}(\omega)$ as a function of $\omega$ of the BH molecule  obtained by a 
[2,1] Pad\'{e} resummation of $\bm{\Sigma}^{(n)}(\omega)$ over $1 \leq n \leq 5$. The corresponding elements of 
the exact $\bm{\epsilon} + \bm{\Sigma}(\omega)$ and $\bm{\epsilon} + \bm{\Sigma}^{(5)}(\omega)$ are also shown.}
\label{fig:BH_Pade21}
\end{figure}

\begin{figure}
  \includegraphics[scale=0.65]{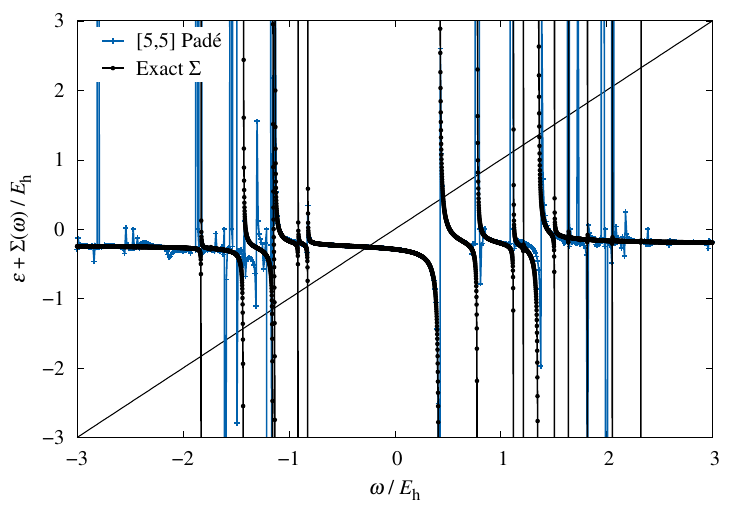}
\caption{The same as Fig.\ \ref{fig:BH_Pade21} but for a [5,5] Pad\'{e} resummation of $\bm{\Sigma}^{(n)}(\omega)$ over $1 \leq n \leq 12$. }
\label{fig:BH_Pade55}
\end{figure}

Figures \ref{fig:BH_Pade21} and \ref{fig:BH_Pade55} show the [2,1] and [5,5] Pad\'{e} approximants of the perturbative self-energies of the BH molecule. 
They were obtained by using Eqs.\ (12)--(17) of Laidig and Bartlett \cite{laidig} of which their ``$E^{(n)}$'' should be read as $\Sigma^{(n)}$ in our case. 
The $[N,N-1]$ and $[N,N]$ Pad\'{e} approximants sum over perturbative self-energies through $2N+1$ and $2N+2$ orders, respectively, at no additional cost. 
The [2,1] Pad\'{e} approximant (Fig.\ \ref{fig:BH_Pade21}) not only largely eliminates the incorrect (concave and convex) shapes of $\Sigma^{(5)}$,
but it also shifts the positions of the singularities in the right directions to be in better agreement with the exact self-energy. The [5,5] Pad\'{e} approximant  (Fig.\ \ref{fig:BH_Pade55}) traces 
the exact self-energy even more closely, although spurious spikes can be seen. 

In this context, it may be instructive to compare the $\Delta$MP$n$ method \cite{pickup,deltamp,Hirata2017,Smiga2018} with MBGF($n$).
 $\Delta$MP$n$ computes the perturbative self-energy as the difference in the $n$th-order many-body perturbation energy 
between the frozen-orbital $N\pm 1$ and $N$ electron systems. At the second and third orders, the $\Delta$MP$n$
self-energy agrees with the MBGF($n$) self-energy in the diagonal, frequency-independent approximation \cite{deltamp,Hirata2017}. At fourth and higher orders, these two self-energies deviate from each other, but they both converge at the same exact 
self-energy at an infinite order (unless they diverge). $\Delta$MP$n$ can reach the exact limit by including semi-reducible and linked-disconnected diagrams,
 which correct the errors from the diagonal and frequency-independent approximations, respectively. These strange diagrams (also implicit 
 in EOM-CC)  are, however, 
illegal in MBGF($n$) \cite{deltamp,Hirata2017}. 

Unlike MBGF($n$), which is nonconvergent for most roots, $\Delta$MP$n$ is only occasionally divergent 
(see Fig.~6 of Ref.\ \cite{Hirata2017}) as it is based on the Hirschfelder--Certain degenerate perturbation theory (HCPT) \cite{hirschfelder}
applied to individual states. This again underscores the fact that the positions of the poles can be directly and more reliably expanded 
in converging Taylor series; only when they are placed in the denominator of a rational function does a perturbation theory struggle to 
shift their positions in a systematic, converging manner. 
A Pad\'{e} resummation of the perturbative self-energies works well because it expands 
the numerator and denominator of a Green's function separately
and the power-series expansion of the denominator is analogous to $\Delta$MP$n$. 

\section{Conclusions}

The Feynman--Dyson diagrammatic perturbation expansions of the self-energy 
and Green's function are
reliably convergent and thus physically sound only in some small domains of $\omega$, 
which includes the central overlapping bracket around $\omega=0$ enclosing most principal roots 
and some low-energy satellite roots. 
Outside these domains, 
the perturbation expansions are nonconvergent. 
Our mathematical analysis suggests that the convergence is assured only in an overlap of the exact and mean-field $\omega$ brackets.  
An odd-order MBGF($n$) method can have roots that are complex or whose
residues are outside the valid range of zero to one, whereas
 a higher-even-order self-energy  tends to consist of many vertical lines, predicting phantom poles with zero residues. 
 A majority of satellite roots of MBGF($n$) are distributed in nearly the same frequency domains spanned by mere HF orbital energy differences and include 
 little to no electron-correlation effects, no matter how high the perturbation order is raised.

This is in contrast with the $\Delta$MP$n$ method \cite{pickup,deltamp,Hirata2017,Smiga2018}, in which 
IPs and EAs are computed as the energy differences of the $n$th-order HCPT \cite{hirschfelder} applied 
to the frozen-orbital $N\pm1$ and $N$ electron systems. The HCPT energies for both principal and satellite ionized or
electron-attached states \cite{hirschfelder} are only occasionally divergent, and when they converge, they do so at the exact energies. 
Even though both $\Delta$MP$n$ and MBGF($n$) can be formulated in terms of the HCPT corrections to the identical zeroth-order wave functions and energies \cite{Hirata2017}, 
the former is mostly sound for both principal and satellite states, while the latter is pathological for most satellite states. 

Infinite partial resummations of diagrams can exacerbate the nonconvergence. 
The summation of the ring and ladder diagrams up to an infinite order by vertex renormalization (TDA) deteriorates 
the self-energy outside the central bracket. Worse still, the result of calculation changes  dramatically and alternately with 
the number of cycles taken to solve the amplitude equations in an iterative algorithm, making the method ill-defined when all roots are sought.
The summation of all tower diagrams by edge renormalization (self-consistent Green's function method) has a factorially increasing
number of roots, whose energies encroach on the frequency domains where no exact roots can be found. 

This is in sharp contrast with coupled-cluster theory. 
TDA is to IPs and EAs as LCCD is to  ground-state energies since they both sum over the same infinite set of ring and ladder diagrams. 
Despite this similarity, TDA is pathological for low- and high-lying satellite roots, while
coupled-cluster theory  via the EOM-CC formalism offers the most accurate, robust, and converging 
approximations for all roots obtained directly as eigenvalues of a matrix
\cite{stanton_eom,Stanton_ip,Nooijen_ea2,Bartlett_ip,Stanton_ip2,Gour2005,Kamiya_ip,Gour2006,Kamiya_ea,hirata_ipeomcc,Bartlett2012}. 
The EOM formulation of MBGF \cite{Rowe1968,simons_3rd,herman,Herman1980,Herman1980_2}, which can also be viewed as an infinite partial resummation of 
diagrams, 
is recast into a matrix diagonalization \cite{BakerPickup} and may thus be free from the convergence difficulties faced by MBGF($n$). 

The nonconvergence of  the Feynman--Dyson perturbation theory, which is the mathematical foundation of quantum field theory \cite{Feynman,dyson2,DysonS1949,schwinger,sakurai,dyson_physicsworld},  makes it 
hard to use higher-than-second-order self-energy and Green's function in the ans\"{a}tze that are predicated on the knowledge of all poles and residues.
Such ans\"{a}tze include the Galitskii--Migdal identity \cite{Galitskii1958,Koltun1972,Holleboom1990,Ortiz_contour1995}, self-consistent Green's function methods \cite{luttingerward,BaymKadanoff1961,Baym_selfconsistent,VanNeck1991,Dickhoff_chapter7,Dickhoff2004,Dahlen2005,Barbieri2006,Barbieri2009,Phillips2014,Neuhauser2017,Tarantino2017,CoveneyTew2023}, 
and some models of the ADC \cite{schirmer1982,schirmer} that evaluate the static part of its self-energy by summing over all poles of a Feynman--Dyson perturbative self-energy
(see Deleuze {\it et al.}\ \cite{Deleuze1995,Deleuze_sizeconsistency} for the ADC's lack of size- or charge-consistency, which may be related to the pathologies discussed here). 
Insofar as the self-consistent Green's function methods are derived from 
the diagrammatic Luttinger--Ward functional \cite{luttingerward,BaymKadanoff1961,Baym_selfconsistent,Kozik2015,Rossi2015,Gunnarsson2017,Lin2018}, on which the formalisms of DMFT \cite{Georges1996} are also based,
these methods may be adversely affected by the ill-defined nature of the perturbative Feynman propagators at many frequencies. 

The nonconvergence is rooted in the rational-function form of the exact Green's function,  
featuring numerous singularities of the forms $(\omega-E_0 + E_I)^{-1}$ and $(\omega-E_A + E_0)^{-1}$ even before a perturbation approximation is introduced. 
This function is nonanalytic at many frequencies and thus cannot be expanded in converging power series, which is ultimately why perturbation theory tends to fail. 
This is reminiscent of the  
Bardeen--Cooper--Schrieffer theory
of superconductivity  \cite{Bardeen2}, whose superconducting gap formula, $2\delta\, e^{-1/\rho V}$, is nonanalytic and cannot be expanded in a converging power series
of the electron-phonon coupling $V$, even though it is small. To such a problem, ``perturbation theory would not be easy to apply'' (in page 224 of Ref.\ \cite{march}). 
Likewise, the Kohn--Luttinger nonconvergence \cite{kohn,luttingerward} of finite-temperature many-body perturbation theory \cite{JhaHirata,HirataJha,HirataJha2,JhaHirata_canonical,HirataJCP} is ascribed to the nonanalyticity of the rational-function forms of the exact 
grand potential and internal energy at $T=0$  \cite{hiratapra,HirataCPL}.

A Pad\'{e} approximant is a power-series expansion of a rational function and is particularly promising for approximating a Feynman propagator, 
which is a nonanalytic rational function. Pad\'{e} approximants were shown to accelerate the convergence of many-body perturbation energies, 
sometimes transforming a divergent series into a rapidly convergent one at no additional cost \cite{Goscinski1967,Brandas1970,GoscinskiBrandas1971,Bartlett1972,Bartlett1973,laidig,hirata_cc}. 
For MBGF($n$), Pad\'{e} approximants exhibit remarkable performance, largely restoring the correct functional form of the self-energy and shift its poles 
in the right directions by the right amounts, which is consistent with our identified cause of the nonconvergence.


\acknowledgments
We thank Professor Piotr Piecuch (Michigan State University), Dr.\ Xiuyi Qin (CGG), Professor Takashi Nakatsukasa (University of Tsukuba), 
and Professor Vojtech Vl\v{c}ek (University of California, Santa Barbara) for insightful discussions and constructive suggestions.

S.H.\ was supported by the U.S. Department of Energy (DoE), Office of Science, Office of Basic Energy Sciences under Grant No.\ DE-SC0006028 and also by the Center for Scalable Predictive methods for Excitations and Correlated phenomena (SPEC), which is funded by the U.S. DoE, Office of Science, Office of Basic Energy Sciences, Division of Chemical Sciences, Geosciences and Biosciences as part of the Computational Chemical Sciences (CCS) program at Pacific Northwest National Laboratory (PNNL) under FWP 70942. PNNL is a multi-program national laboratory operated by Battelle Memorial Institute for the U.S. DoE.
S.H.\ is a Guggenheim Fellow of the John Simon Guggenheim Memorial Foundation. 

I.G.\ was supported by Polish National Science Center under Grant No.\ 2020/39/O/ST4/00005

R.J.B.\ was supported by the Air Force Office of Scientific Research under AFOSR Award No.\ FA9550-19-1-0091.

\bibliography{library.bib}

\end{document}